\documentclass[12pt, reqno]{amsart}

\allowdisplaybreaks

\usepackage[all]{xy}
\usepackage{amssymb} 

\newcommand{\morespace}{}

\newcommand{\wt}{\widetilde}
\newcommand{\wh}{\widehat}

\begin{document}
\morespace

\title[Microlagrangian manifolds and fluctuations]{
Microlagrangian manifolds and\\
quasithermodynamic fluctuations\\ 
of nonequilibrium states
}

\author[A. E. Ruuge]{Artur E. Ruuge}

\address{
Department of Mathematics and Computer Science, 
University of Antwerp, 
Middelheim Campus Building G, 
Middelheimlaan 1, B-2020, 
Antwerp, Belgium
}
\email{artur.ruuge@ua.ac.be
}


\begin{abstract}
\morespace

The paper deals with ``quantization'' and ``second quantization'' of phenomenological thermodynamics 
with respect to the Boltzmann's constant. 
It is suggested to perceive the quasithermodynamic parameter (corresponding to the Boltzmann's constant) 
as a mathematical analogue of the semiclassical parameter (corresponding to the Planck's constant), and to  
introduce a new concept of a ``thermocorpuscle'' (a thermodynamic analogue of a particle where 
the coordinates are replaced by the nonequilibrium thermodynamic forces and 
the momenta are replaced by the corresponding flows). 
The semiclassical quantization of phenomenological thermodynamic Lagrangian manifolds 
yields a new system of equations for the quasithermodynamic fluctuations along 
a curve of evolution of a nonequilibrium physical system. 
This leads to 
a quasithermodynamic analogue of Bell's inequalities 
and their violation is a new effect that can be tested experimentally.   
The generating function of the quasithermodynamic fluctuations 
(the nonequilibrium analogue of a partition function) is interpreted 
as an expectation value of a second quantized operator expressed 
via the density of ``thermocorpuscles''. 
An analogue of the BBGKY chain of equations defines a 
deformation of the fluctuations by an interaction between the thermocorpuscles. 
In place of an interaction parameter in mechanics (the ``external'' Planck's constant), 
one introduces the ``external'' Boltzmann's constant for an asymptotic expansion of the thermodynamic collision integral. 

\end{abstract}

\maketitle

\section{Introduction}

The original motivation for this paper stems from an idea 
to study the quasithermodynamic fluctuations in nonequilibrium statistical physics 
using the methods of semiclassical approximation of quantum theory. 
If we look at a semiclassical wave function $\psi_{\hbar} (x)$, 
of a quantum system with $n$ degrees of freedom $x = (x_1, x_2, \dots, x_n)$, 
and take a classical limit $\hbar \to 0$ for the corresponding Wigner's quasiprobability function 
$\rho [\psi_{\hbar}] (x, p)$, $p = (p_1, p_2, \dots, p_n)$, 
then under some mild conditions and assumptions \cite{Karasev_Maslov}, we obtain a smooth manifold 
\begin{equation*}
\Lambda := \mathrm{supp} \lim_{\hbar \to 0} \rho [\psi_{\hbar}] (x, p), 
\end{equation*}
where the limit on the right-hand side is understood in the weak sense. 
This manifold is a Lagrangian submanifold $\Lambda \subset \mathbb{R}_{x, p}^{2 n}$ in the 
classical phase space $\mathbb{R}_{x, p}^{2 n}$ with respect to the canonical 
symplectic structure $\omega = \sum_{i = 1}^{n} d p_i \wedge d x_i$. 

The Lagrangian manifolds arise in a perfectly natural way in the 
equilibrium thermodynamics as well. 
Consider, for example, 
a physical system consisting of $\nu$ moles of a chemical substance in a closed volume $V$.  
Then what happens in the molecular kinetic theory is that one splits $\nu$ into a product 
of a ``very small'' quantity $k_B$ (the Boltzmann's constant) and a ``very big'' quantity $N$ 
(the number of particles in the system): 
\begin{equation} 
\label{eq:nu_kB_N}
\nu = \frac{1}{R} k_B N, 
\end{equation}
where $R$ is a constant fixing the units of measurement (the universal gas constant). 
The basic idea is to consider a family of underlying systems with $N_{\lambda} = \lambda N$ particles 
and volume $V_{\lambda} = \lambda V$, where $\lambda$ is the \emph{rescaling parameter}, and to 
derive the empirical laws from the \emph{thermodynamic limit} $\lambda \to \infty$.  
As a model example, take the underlying system 
to be a system of $N_{\lambda}$ particles on a $n$-dimensional torus of radius $L_{\lambda} = \lambda^{1/ n} L$, $V = L^n$, $L$ is fixed. 
Assume that it is described by an 
energy spectrum $E_{0}^{(\lambda)} < E_{1}^{(\lambda)} < \dots < E_{m}^{(\lambda)} < \dots$, 
where each level $E_{m}^{(\lambda)}$, $m \in \mathbb{Z}_{\geqslant 0}$, 
has a finite degree of degeneracy $g_{m}^{(\lambda)}$. 
One can  
construct a \emph{partition function} (assume that the power series converges): 
\begin{equation*} 
\mathcal{Z}_{k_B}^{(\lambda)} (\beta) := 
\sum_{m = 0}^{\infty} g_{m}^{(\lambda)} \exp \Big(- \frac{1}{k_B} \beta E_{m}^{(\lambda)} \Big), 
\end{equation*}
where $\beta = T^{-1}$ is the inverse absolute temperature of the system. 
The \emph{free energy} $\mathcal{F}_{k_B}^{(\lambda)} (T)$ of the system at temperature $T$ is given by 
\begin{equation*} 
\mathcal{F}_{k_B}^{(\lambda)} (T) := - k_B T \mathrm{ln} \mathcal{Z}_{k_B}^{(\lambda)} (T^{-1}). 
\end{equation*}
Take some values of $\nu$ and $V = L^n$, and 
assume that the free energy 
satisfies 
\begin{equation*} 
\lambda^{-1} \mathcal{F}_{k_B}^{(\lambda)} (T) = \nu f_{k_B} (T, v) + o (\lambda^{-1}), 
\end{equation*}
as $\lambda \to \infty$, where $v := V / \nu$ is the volume per mole, and 
$f_{k_B} (T, v)$ is a smooth function. 
In the limit $\lambda \to \infty$ we have: 
\begin{equation} 
\label{eq:entropy_pressure}
s = - \frac{\partial f_{k_B} (T, v)}{\partial T}, \quad 
p = - \frac{\partial f_{k_B} (T, v)}{\partial v}, 
\end{equation}
where $s = S/ \nu$, $S$ is the \emph{entropy} of the system, and $p$ is the \emph{pressure} in the system. 
This is precisely what one can see in the \emph{phenomenological} thermodynamics. 
The equations \eqref{eq:entropy_pressure} define a Lagrangian manifold $L \subset \mathbb{R}^{4} (T, v, s, p)$ 
with respect to the symplectic structure 
\begin{equation*}
\omega := d s \wedge d T - d v \wedge d p.
\end{equation*}
According to \cite{Maslov_thermo1,Maslov_thermo2,Maslov_thermo3}, 
the \emph{intensive} thermodynamic coordinates $(T, - p)$ can be perceived as ``coordinates'', and 
the \emph{extensive} thermodynamic coordinates $(s, v)$ can be perceived as ``momenta''. 
More generally,  
one may perceive an abstract thermodynamic system with $d$ degrees of freedom 
$\xi = (\xi_1, \xi_2, \dots, \xi_d)$ as a Lagrangian submanifold $\Lambda \subset 
\mathbb{R}_{\xi, \eta}^{2 d}$ with respect to symplectic structure 
$\omega = \sum_{i = 1}^{d} d \eta_i \wedge d \xi_i$, 
$\eta = (\eta_1, \eta_2, \dots, \eta_d)$. 
The Lagrangian manifolds are a central object in the semiclassical approximation of quantum mechanics 
\cite{Maslov_asymp_meth, Maslov_op_meth} and it is of interest 
to investigate the corresponding analogy between thermodynamics and mechanics in more detail. 
Recently this topic has 
attracted some additional attention and has received some new interesting developments 
in \cite{Coutant_Rajeev,Maslov_ultra,Maslov_Nazaikinskii,Rajeev1,Rajeev2}.  

In the present paper we prefer to use a little different coordinates to describe 
the thermodynamic Lagrangian manifolds. 
Denote $\varepsilon = E/ \nu$ 
the \emph{internal energy} of the system per mole. 
Then at temperature $T = \beta^{-1}$ we have: 
\begin{equation*} 
\varepsilon = \frac{\partial}{\partial \beta} (\beta f_{k_B} (\beta, v)).  
\end{equation*} 
The first law of thermodynamics tells us: 
\begin{equation*} 
d s (\varepsilon, v) = \beta d \varepsilon + \wt{p} d v, 
\end{equation*}
where $\wt{p} := p/ T$, and the entropy per mole is written as a function of $\varepsilon$ and $v$, 
$s = s (\varepsilon, v)$. 
The derivatives $\beta = (\partial s/ \partial \varepsilon)_{v}$ and 
$\wt{p} = (\partial s/ \partial \beta)_{\varepsilon}$ define 
a Lagrangian manifold $\Lambda_{\mathit{thermo}} \subset \mathbb{R}^{4} (\varepsilon, v, \beta, \wt{p})$ 
with respect to the symplectic structure 
\begin{equation*}
\omega_{\mathit{thermo}} := d \beta \wedge d \varepsilon + d \wt{p} \wedge d v. 
\end{equation*}
Therefore, conceptually, the entropy (per mole) of a system is an \emph{action} on $\Lambda_{\mathit{thermo}}$. 
The total entropy $S_{\mathit{thermo}} := \nu s$ is measured in the same units as the Boltzmann constant: 
\begin{equation} 
\label{eq:S_thermo_kB}
[S_{\mathit{thermo}}] = [k_B]. 
\end{equation} 
If we have an abstract mechanical system with coordinates $Q = (Q_1, Q_2, \dots, Q_n)$ and 
momenta $P = (P_1, P_2, \dots, P_n)$ and consider a Lagrangian manifold 
$\Lambda_{\mathit{mech}}$ in the phase space $(\mathbb{R}_{Q, P}^{2 n}, \omega_{\mathit{mech}})$, 
where 
\begin{equation*}
\omega_{\mathit{mech}} := \sum_{i = 1}^{n} d P_{i} \wedge d Q_{i}, 
\end{equation*}
then we have an 
action $S_{\mathit{mech}} := \int_{\gamma} P d Q$, $\gamma$ is smooth curve on $\Lambda_{\mathit{mech}}$, 
which is measured in the same units as the Planck's constant: 
\begin{equation} 
\label{eq:S_mech_hbar}
[S_{\mathit{mech}}] = [\hbar]. 
\end{equation}
In this context, the \emph{Heisenberg's uncertainty principle} is totally similar to the 
\emph{Einstein's formula} for the quasithermodynamic fluctuations: 
\begin{equation} 
\label{eq:uncertainty}
\Delta P_{i} \Delta Q_j \sim \hbar, \quad 
\Delta E_{l} \Delta \beta_{m} \sim k_B, 
\end{equation}
where $\Delta$ denotes the standard deviation describing a fluctuation, 
$i, j = 1, 2, \dots, n$, and $l, m = 1, 2$, 
and we put 
$E_{1} = E$, $E_{2} = V$, $\beta_{1} = \beta$, $\beta_{2} = \wt{p}$. 
Informally speaking, $\hbar$ is a quantum of mechanical action, and 
$k_{B}$ is a ``quantum'' of entropy. 

The analogy 
between the Planck's constant $\hbar$ and the Boltzmann's constant $k_B$ expressed, 
in particular, by 
\eqref{eq:S_thermo_kB},\eqref{eq:S_mech_hbar},\eqref{eq:uncertainty}, 
is the central motive of the present paper. 
The general scheme is as follows. 
First, we generalize the idea of quantization of phenomenological thermodynamic Lagrangian manifolds 
to the nonequilibrium setting.  
This defines a generic shape of the equations describing the transfer of 
fluctuations along a nonequilibrium evolution curve. 
Under some assumptions one can assemble these fluctuations into a ``quasithermodynamic wavefunction'' 
and extract the thermodynamic Hamilton-Jacobi equation in the limit $\lambda \to \infty$. 
It turns out, that in \emph{nonequilibrium} quasithermodynamics 
one can mimic Bell's inequalities and 
it is natural to perceive the thermodynamic forces $X = (X_1, X_2, \dots, X_d)$ as 
``coordinates'', and the corresponding flows $J = (J_1, J_2, \dots, J_d)$ 
as ``momenta''. 
We apply the second quantization to produce the ``thermocorpuscles'' 
introducing the creation and annihilation operators $a^{\pm} (X, J)$ in 
a phase space point $(X, J) \in \mathbb{R}_{X, J}^{2 d}$, 
the symplectic structure is $\Omega := \sum_{i = 1}^{d} d J_{i} \wedge d X_{i}$.  
Basically, the thermodynamic analogue of the Wigner's equation for 
the single particle quasiprobability distribution  
is lifted to the symmetric Fock space $\mathcal{F}^{\#}$, 
\begin{equation*} 
\mathcal{F}^{\#} := \mathbb{C} \oplus L^2 (\mathbb{R}_{X, J}^{2 d} ) \oplus 
(L^2 (\mathbb{R}_{X, J}^{2 d} ))^{\otimes_{\mathit{symm}} 2} \oplus \dots, 
\end{equation*}
where $\otimes_{\mathit{symm}}$ denotes the symmetric tensor product. 
After that, it is suggested to deform the second 
quantized equations by an interaction between the thermocorpuscles and to mimic the basic 
constructions of statistical mechanics like the BBGKY chain and the collision integral.   
In particular, this yields a new interpretation of the nonequilibrium analogue 
$\mathcal{Z}_{k_B}^{(\lambda)} (u; t)$ of the 
partition function in terms of the density of thermocorpuscles, 
\begin{equation*} 
\mathcal{Z}_{k_B}^{(\lambda)} (u; t) = \int d X \, 
\exp(- u X/ k_{B})
\int d J \, 
\langle R_{\lambda}^{t}, 
a^{+} (X, J) a^{-} (X, J) 
R_{\lambda}^{t} \rangle, 
\end{equation*} 
where $u = (u_1, u_2, \dots, u_d)$ 
is a parameter varying in a neighbourhood of $\bar 0 = (0, 0, \dots, 0)$, 
$R_{\lambda}^{t} \in \mathcal{F}^{\#}$ is a real vector depending on the rescaling parameter 
$\lambda$ and the time $t$, and $\langle -, - \rangle$ denotes the inner product.

\section{Equilibrium quasithermodynamics}

Let us look at a phenomenological one-component thermodynamic system with two degrees of freedom, $d = 2$. 
To be more concrete, let the molar extensive coordinates be $\xi_1 = \varepsilon$ and $\xi_2 = v$, 
where $\varepsilon := E/ \nu$, $v := V/ \nu$,  
$E$ is the internal energy, $V$ is the volume of the system, $\nu$ is the number of moles. 
We have a phase space $\mathbb{R}^{4} (\xi_1, \xi_2, \eta_2, \eta_2)$ with a symplectic structure 
$\omega = d \eta_1 \wedge d \xi_1 + d \eta_2 \wedge d \xi_2$, 
where $\eta_1 = \beta$ is the inverse absolute temperature of the system, and 
$\eta_2 = \wt{p} := \beta p$, $p$ is the pressure in the system. 
The system is described  as a two-dimensional Lagrangian manifold $\Lambda$ in this phase space 
defined by equations  
\begin{equation*} 
\beta = \Big( \frac{\partial s}{\partial \varepsilon} \Big)_{v}, \quad 
\wt{p} = \Big( \frac{\partial s}{\partial v} \Big)_{\varepsilon}, 
\end{equation*}
where $s = S/ \nu$ is the entropy $S$ of the state of the system taken per mole. 
The function $s = s (\varepsilon, v)$ is a generating function of $\Lambda$ with respect to the 
focal coordinates $(\varepsilon, v)$. If we assume, that $\Lambda$ admits another choice of focal coordinates, 
say, $(\beta, \wt{p})$, then the corresponding generating function 
$\wt{\mu} = \wt{\mu} (\beta, \wt{p})$ is linked to $s = s (\varepsilon, v)$ via the \emph{Legendre transform}, 
\begin{equation*} 
\wt{\mu} (\beta, \wt{p}) \big|_{\Lambda} = \big[ 
s (\varepsilon, v) - \beta \varepsilon - \wt{p} v 
\big] \big|_{\Lambda}. 
\end{equation*}
The physical meaning of this function is just $\wt{\mu} = - \beta \mu$, 
where $\mu$ is the \emph{chemical potential} of the system. 
It is natural to consider along with $\Lambda$ a Lagrangian manifold 
$\Lambda^{+}$ of dimension $d + 1 = 3$ in the phase space 
$\mathbb{R}^{6} (x, y)$ with coordinates 
$x = (x_1, x_2, x_3)$, $x_1 = E$, $x_2 = V$, $x_3 = \nu$, and 
``momenta'' $y = (y_1, y_2, y_3)$, $y_1 = \beta$, $y_2 = \wt{p}$, $y_3 = \wt{\mu}$. 
The symplectic structure $\omega^{+}$ is given by 
$\omega^{+} := \sum_{i = 1}^{3} d y_i \wedge d x_i$, 
and the manifold $\Lambda^{+}$ can be described as 
\begin{equation*} 
\Lambda^{+} := \lbrace (x, y) \,|\, x_3 \not = 0, \, 
(x_1/ x_3, x_2/ x_3, y_1, y_2) \in \Lambda, \, 
y_3 = \wt{\mu} (x_1/ x_3, x_2/ x_3) \rbrace. 
\end{equation*} 
One can observe, that $\Lambda^{+}$ admits a projectivization with respect to the extensive 
coordinates $x = (x_1, x_2, x_3) = (E, V, \nu)$, which can be taken as global focal coordinates on $\Lambda^{+}$. 
At the same time, all the points of $\Lambda^{+}$ are singular 
with respect to the focal coordinate plane $\mathbb{R}^3 (y)$, $y = (y_1, y_2, y_3) = (\beta, \wt{p}, \wt{\mu})$. 
Our physical system can be identified with this $\Lambda^{+}$, and 
one should assume (for the physical reasons), that $\Lambda^{+}$ is connected and simply connected 
\cite{Maslov_thermo1, Maslov_Nazaikinskii}. 

It can happen, that $\Lambda^{+}$ admits other focal charts 
with a choice of coordinates different from $x = (x_1, x_2, x_3)$. 
Recall the notation: 
\begin{equation*} 
(x_1, x_2, x_3) = (E, V, \nu), \quad (y_1, y_2, y_3) = (\beta, + \beta p, - \beta \mu). 
\end{equation*}
In a nonsingular chart $U$ with coordinates $(x_1, x_2, x_3) = (E, V, \nu)$, 
the generating function $S (x_1, x_2, x_3)$ is the entropy of the system 
in a point $\alpha \in U \subset \Lambda^{+}$ corresponding to $(x_1, x_2, x_3)$, 
$(d S - \sum_{i = 1}^{3} y_i d x_i)|_{\Lambda^{+}} = 0$, and 
$S (E, V, \nu) = \nu s (\varepsilon, v)$, $\varepsilon = E/ \nu$, $v = V/ \nu$. 
In a singular chart $U_1$ with coordinates $(y_1, x_2, x_3) = (\beta, V, \nu)$ the generating 
function $S_1 (y_1, x_2, x_3)$ can be interpreted (up to an additive constant) 
as $- \beta F$, where $F$ is the free energy of the system in a point $\alpha \in U_1 \subset \Lambda^{+}$ 
corresponding to $(y_1, x_2, x_3)$. 
In a singular chart $U_{1, 3}$ with coordinates $(y_1, x_2, y_3) = (\beta, V, - \beta \mu)$, 
the generating function $S_{1, 3} (y_1, x_2, y_3)$ can be interpreted (up to an additive constant) 
as $- \beta \Omega$, where $\Omega$ is the thermodynamic potential with respect to $(T, V, \mu)$
taken in a point $\alpha \in U_{1, 3} \subset \Lambda^{+}$ corresponding to $(y_1, x_2, y_3)$. 

Let us analyse the case of the to charts $U_1$ and $U_{1, 3}$ in more detail. 
Assume $U_1 \cap U_{1, 3} \not = \emptyset$ and fix $\alpha \in U_1 \cap U_{1, 3}$. 
In the ambient space $\mathbb{R}^{6} (x, y) \supset \Lambda^{+}$ this point 
acquires coordinates $x (\alpha) = (x_1 (\alpha), x_2 (\alpha), x_3 (\alpha))$ and 
$y (\alpha) = (y_1 (\alpha), y_2 (\alpha), y_{3} (\alpha))$. 
The step from \emph{phenomenological} thermodynamics to \emph{statistical} thermodynamics 
corresponding to the first chart consists in the following. 
One considers a family of physical systems parametrized by $\lambda > 0$ 
(the \emph{rescaling parameter}), $\lambda \to \infty$. 
Every system is placed in a thermostat and the walls of the system are fixed and are impenetrable for the particles. 
The inverse temperature $\beta_{\lambda} (\alpha) = x_1 (\alpha)$ is the same for each $\lambda$, 
the number of particles $N_{\lambda} (\alpha) = \lambda x_3 (\alpha) R/ k_B$, together with the volume of the system 
$V_{\lambda} (\alpha) = \lambda x_2 (\alpha)$, grow  linearly as $\lambda \to \infty$.  
For simplicity, one may have in mind a system of $N$ particles moving on a $n$-dimensional torus 
of radius $L$ described by a Hamiltonian $\wh{H}_{N}^{(L)} (g)$, where $g$ is a collection of 
parameters describing the interaction between the particles and the geometrical shape of the volume 
(the external potential). 
One substitutes $L = (V_{\lambda} (\alpha))^{1/ n}$ and $N = N_{\lambda} (\alpha)$, assuming that 
the parameters $g = g_{\lambda} (\alpha)$ may be adjusted as well.  
The free energy 
$\mathcal{F}_{k_B, N}^{(L)} (T; g)$ at absolute temperature $T$ is given by 
$\mathcal{F}_{k_B, N}^{(L)} (T; g) = - k_{B} T \mathrm{ln} \mathcal{Z}_{k_B, N}^{(L)} (T^{-1}; g)$, 
where 
\begin{equation*} 
\mathcal{Z}_{k_B, N}^{(L)} (\beta; g) := 
\mathrm{Tr} \exp \Big( - \frac{1}{k_B} \beta \wh{H}_{N}^{(L)} (g) \Big), 
\end{equation*}
is the \emph{partition function} of the canonical Gibbs distribution at absolute temperature 
$T = \beta^{-1}$
(we assume that the corresponding trace is finite 
for the required values of $\beta$, $L$, $N$, and $g$). 
The link with phenomenological thermodynamics is as follows: 
\begin{equation} 
\label{eq:phenomen_link_S1}
S_{1} (y_1 (\alpha), x_3 (\alpha), x_3 (\alpha)) = 
\lim_{\lambda \to \infty} \big[ - \lambda^{-1} \beta_{\lambda} (\alpha) 
\mathcal{F}_{k_B, N_{\lambda} (\alpha)}^{((V_{\lambda} (\alpha))^{1/ n})} (\beta_{\lambda} (\alpha)^{-1}; g_{\lambda} (\alpha)) + C (\lambda) \big],   
\end{equation}
where it is assumed that 
one can find functions $g_{\lambda} (\alpha)$ and $C (\lambda)$ so that 
the limit on the right-hand side exists. 
Denote 
$S_{1}^{(\lambda)} (\alpha)$ 
the expression in the square brackets on the right-hand side of \eqref{eq:phenomen_link_S1}.

Now if we take the other chart $U_{1, 3}$ with coordinates $(y_1, x_2, y_3) = (\beta, V, - \beta \mu)$, 
then we need to consider \emph{another family} of rescaled systems of \emph{another sort}: 
the walls of the system must admit a penetration of particles, i.e. 
the border of the system is described purely geometrically. 
One still has a rescaling parameter $\lambda > 0$, 
the volume of the rescaled system $V_{\lambda} (\alpha) = \lambda x_2 (\alpha)$, 
but now one should look at 
the partition function 
$\zeta_{k_B}^{(L)} (\beta, \mu; g)$ 
of the \emph{grand canonical ensemble}, 
\begin{equation*} 
\zeta_{k_B}^{(L)} (\beta, \mu; g) := 
\mathrm{Tr} \exp \Big( 
-\frac{1}{k_B} \beta \Big[ 
\wh{\mathcal{H}}^{(L)} (g) - 
\mu \wh{\mathcal{N}}^{(L)} 
\Big]
\Big), 
\end{equation*}
(it is assumed that the trace is finite for the required values of $\beta$, $\mu$, and $g$), 
where $\beta$ and $\mu$ are 
the inverse absolute temperature and the chemical potential 
determined by the environment of the system, respectively, 
$\wh{\mathcal{H}}^{(L)} (g)$ is the second quantized Hamiltonian of the multiparticle system 
on a $n$-dimensional torus of radius $L$, 
$g$ is a collection of parameters describing the interaction between the 
particles and the external potential, 
$\wh{\mathcal{N}}^{(L)}$ is the second quantized operator of the number of particles in the system \cite{Berezin}. 
Denote 
$\Omega_{k_B}^{(L)} (T, \mu; g) := - k_B T \mathrm{ln} \zeta_{k_B}^{(L)} (T^{-1}, \mu; g)$ 
(the \emph{potential $\Omega$}). 
One needs to substitute in place of $\beta$, $\mu$, and $g$ some 
functions $\beta_{\lambda} (\alpha)$, $\mu_{\lambda} (\alpha)$, and $g_{\lambda} (\alpha)$, respectively, 
in such a way, that 
the average energy and the average number of particles in the system grow linearly with $\lambda \to \infty$. 
It is convenient to 
change the variables 
$y_1 := \beta$, $y_2 := - \beta \mu$, 
and to consider 
$\wt{\mathcal{Z}}_{k_B}^{(L)} (y_1, y_2; g) := 
\zeta_{k_B}^{(L)} (\beta, \mu; g)$. 
Denote $(\beta_{\lambda} (\alpha; g), - \beta_{\lambda} (\alpha; g) \mu_{\lambda} (\alpha; g))$ 
the solution of the system of equations 
\begin{equation*} 
\begin{gathered}
\big( \mathcal{\wt{\mathcal{Z}}}_{k_B}^{((\lambda x_2 (\alpha))^{1/ n})} (y_1, y_2; g) \big)^{-1} 
\Big( - k_B \frac{\partial}{\partial y_1} \Big) 
\mathcal{\wt{\mathcal{Z}}}_{k_B}^{((\lambda x_2 (\alpha))^{1/ n})} (y_1, y_2; g) 
 = \lambda x_1 (\alpha), \\
\big( \mathcal{\wt{\mathcal{Z}}}_{k_B}^{((\lambda x_2 (\alpha))^{1/ n})} (y_1, y_2; g) \big)^{-1} 
\Big( - k_B \frac{\partial}{\partial y_2} \Big) 
\mathcal{\wt{\mathcal{Z}}}_{k_B}^{((\lambda x_2 (\alpha))^{1/ n})} (y_1, y_2; g) 
 = \lambda x_3 (\alpha) R/ k_B, 
\end{gathered}
\end{equation*}
with respect to $(y_1, y_2)$, assuming it exists and is unique for the required values of $g$. 
Recall, that $R$ denotes the universal gas constant, 
$x_1 (\alpha)$ corresponds to the internal energy, 
$x_2 (\alpha)$ corresponds to the volume, 
and $x_3 (\alpha)$ corresponds to the number of moles of the chemical substance in the phenomenological system. 
The link with the phenomenological thermodynamics is as follows: 
\begin{multline} 
\label{eq:phenomen_link_S13}
S_{1, 3} (y_1 (\alpha), x_2 (\alpha), y_3 (\alpha)) 
= \\ = 
\lim_{\lambda \to \infty} 
\Big[ 
- \lambda^{-1} \beta_{\lambda} (\alpha; g) \Omega_{k_B}^{((\lambda x_2 (\alpha))^{-1})} 
(\beta_{\lambda} (\alpha; g)^{-1}, \mu_{\lambda} (\alpha; g); g)
+ \wt{C} (\lambda) \Big] \Big|_{g = \wt{g}_{\lambda} (\alpha)}, 
\end{multline}
for some functions 
$\wt{g}_{\lambda} (\alpha)$ and $\wt{C} (\lambda)$ ensuring the existence of 
the limit on the right-hand side. 
Denote $S_{1, 3}^{(\lambda)} (\alpha)$ the expression in the square 
brackets on the right-hand side in \eqref{eq:phenomen_link_S13}.

The two functions $S_{1} (y_1, x_2, x_3)$ and $S_{1, 3} (y_1, x_2, y_3)$ are related via the \emph{Legendre transform} 
in the third argument. If we perceive them as functions on the thermodynamic phase space $\mathbb{R}^{6} (x, y)$, 
then we have 
\begin{equation*}
S_{1, 3} (y_1, x_2, y_2) \big|_{\Lambda^{+}} (\alpha) = 
\big[ S_{1} (y_1, x_2, x_3) - y_3 x_3 \big] \big|_{\Lambda^{+}} (\alpha),  
\end{equation*}
for any $\alpha \in U_1 \cap U_{1, 3}$. 
Since $\Lambda^{+}$ admits projectivization with respect to the extensive coordinates 
$x = (x_1, x_2, x_3)$, the generating functions in the focal charts are homogeneous 
with respect to these coordinates: 
\begin{equation*} 
S_{1} (y_1, \rho x_2, \rho x_3) = \rho S_{1} (y_1, x_2, x_3), \quad 
S_{1, 3} (y_1, \rho x_2, y_3) = \rho S_{1, 3} (y_1, x_2, y_3), 
\end{equation*}
for any $\rho > 0$. 

Consider now an abstract phenomenological thermodynamic system 
with extensive coordinates 
$x = (x_1, x_2, \dots, x_{d + 1})$ and 
intensive coordinates $y = (y_1, y_2, \dots, y_{d+ 1})$. 
The corresponding thermodynamic phase space 
is defined as $\mathbb{R}^{2 (d + 1)} (x, y)$ equipped 
with canonical symplectic structure $\omega^{+} = \sum_{i = 1}^{d + 1} d y_i \wedge d x_i$. 
The equilibrium states of the system are identified with a 
Lagrangian manifold $\Lambda^{+} \subset \mathbb{R}^{2 (d + 1)} (x, y)$, 
which is connected, simply connected, contained in the 
region $x_i > 0$, $i = 1, 2, \dots, d + 1$, and admits a projectivization with respect to $x$: 
if $(x (\alpha), y (\alpha))$ are the phase space coordinates corresponding to a point $\alpha \in \Lambda^{+}$, 
then for any $\rho > 0$, the coordinates $(\rho x (\alpha), y (\alpha))$ correspond to another 
point in $\Lambda^{+}$. 
Let $I$ be a subset of $[d + 1] := \lbrace 1, 2, \dots, d + 1 \rbrace$. 
Denote 
\begin{equation} 
\label{eq:notation_x_y_I}
(x, y)_{I} := \big( 
\lbrace \overset{j}{x}_{j} \rbrace_{j \in [d + 1] \backslash I}, 
\lbrace \overset{i}{y}_{i} \rbrace_{i \in I} 
\big), 
\end{equation}
where the indices atop denote the order of appearance of the variables in the list 
read from left to the right, for example, if 
$d = 2$, $I = \lbrace 1, 3 \rbrace$, then $(x, y)_{I} = (y_1, x_2, y_3)$. 
In particular, for the emptyset $I = \emptyset$, we have 
$(x, y)_{\emptyset} = (x_1, x_2, \dots, x_{d + 1})$, and 
for $I = [d + 1]$ we have $(x, y)_{[d + 1]} = (y_1, y_2, \dots, y_{d + 1})$. 
A \emph{focal chart} of type $I$ on $\Lambda^{+}$ is an open set $U \subset \Lambda^{+}$ 
together with a fixed choice of local coordinates of the shape $(x, y)_{I}$. 
For every focal chart of type $I$ over  $U$, 
there exists a function $S_{U, I} ((x, y)_{I})$, such that $U$ is described by the equations: 
\begin{equation*} 
y_{i} = \frac{\partial S_{U, I} ((x, y)_{I})}{\partial x_i}, \quad 
x_{j} = - \frac{\partial S_{U, I} ((x, y)_{I})}{\partial y_j}, 
\end{equation*}
where $i \in I$, $j \in [d + 1] \backslash I$. 
Observe, that for any $\rho > 0$, 
\begin{equation*}
S_{U, I} ((\rho x, y)_{I}) = \rho S_{U, I} ((x, y)_{I}), 
\end{equation*}
whenever both left- and right-hand sides are defined. 
Observe also, that the thermodynamic Lagrangian manifold $\Lambda^{+}$ 
never admits a focal chart of type $[d + 1]$, i.e. 
at least one of the local coordinates in a given focal chart must be extensive. 
In case $U$ is connected, the function $S_{U, I}$ is defined up to an additive constant, 
and for a fixed pair of focal charts $(U_1, I)$ and $(U_2, J)$, such that $U_1 \cap U_2 \not = \emptyset$, 
it is always possible to adjust these constants in such a way that 
\begin{equation*} 
S_{U_2, J} ((x, y)_{J}) = S_{U_1, I} ((x, y)_{I}) - \sum_{j \in J \backslash I} {y_j x_j} + 
\sum_{i \in I \backslash J} y_i x_i, 
\end{equation*}
for every $(x, y)$ corresponding to a point $\alpha \in U_{1} \cap U_{2}$.  
This implies, that $S_{U_1, I}$ and $S_{U_2, J}$ are linked via the 
Legendre transform (denote it $\mathcal{L}_{J, I}^{U_{2} \cap U_{1}}$), 
$S_{U_2, J} = \mathcal{L}_{J, I}^{U_2 \cap U_1} (S_{U_1, I})$, for every $(x, y)$ 
corresponding to a point $\alpha \in U_{1} \cap U_{2}$. 
Observe that $\mathcal{L}_{I, I}^{U} = \mathit{id}$, 
for a focal chart $(U, I)$, where $U$ is connected. 
If $U$ admits focal coordinates of another type $J$, then we have 
$\mathcal{L}_{I, J}^{U} \circ \mathcal{L}_{J, I}^{U} = \mathit{id}$. 
Furthermore, if $U$ admits focal charts of types $I$, $J$, and $K$, then 
it is straightforward to check the cocyclicity condition: 
\begin{equation} 
\label{eq:cocycle_L}
\mathcal{L}_{I, K}^{U} \circ \mathcal{L}_{K, J}^{U} \circ \mathcal{L}_{J, I}^{U} = \mathit{id}. 
\end{equation}
Denote $\wh{\pi}_{I} : \Lambda^{+} \to \mathbb{R}^{d + 1}$ the canonical projection 
$\alpha \mapsto (x (\alpha), y (\alpha))_{I}$, for $I \subset [d + 1]$. 

Return now to the example discussed above, $d = 3$. 
We have defined the functions 
$S_1^{(\lambda)} (\alpha)$ and $S_{1, 3}^{(\lambda)} (\alpha)$
(see the definition right after the equations 
\eqref{eq:phenomen_link_S1} and \eqref{eq:phenomen_link_S13}, respectively), 
such that $S_1 (y_1 (\alpha), x_2 (\alpha), x_3 (\alpha)) = \lim_{\lambda \to \infty} S_{1}^{(\lambda)} (\alpha)$, 
for $\alpha \in U_1$, and 
$S_1 (y_1 (\alpha), x_2 (\alpha), y_3 (\alpha)) = \lim_{\lambda \to \infty} S_{1, 3}^{(\lambda)} (\alpha)$, for 
$\alpha \in U_{1, 3}$. 
What happens in the higher order corrections with respect to $\lambda^{-1} \to 0$? 
In \emph{quasithermodynamics} one assumes that these functions admit the following 
\emph{asymptotic} expansions: 
\begin{equation*} 
S_{1}^{(\lambda)} (\alpha) \simeq \sum_{l = 0}^{\infty} \lambda^{- l} \varphi_{1}^{(l)} (\alpha), \quad 
S_{1, 3}^{(\lambda)} (\alpha) \simeq \sum_{l = 0}^{\infty} \lambda^{- l} \varphi_{1, 3}^{(l)} (\alpha), 
\end{equation*}
where $\varphi_{1}^{(l)} (\alpha)$ and $\varphi_{1, 3}^{(l)} (\alpha)$ are smooth functions, $l = 0, 1, 2, \dots$. 
This can be perceived as a condition on a choice of the function $g_{\lambda} (\alpha)$ present in the definitions. 
It is a little more convenient to work in terms of coordinates, 
introducing the functions $\Phi_{1}^{(l)} (y_1, x_2, x_3)$ and 
$\Phi_{1, 3}^{(l)} (y_1, x_2, y_3)$, defined as follows: 
$\Phi_{1}^{(l)} (y_1 (\alpha), x_2 (\alpha), x_3 (\alpha)) = \varphi_{1}^{(l)} (\alpha)$, $\alpha \in U_1$, and 
$\Phi_{1, 3}^{(l)} (y_1 (\alpha), x_2 (\alpha), y_3 (\alpha')) = \varphi_{1, 3}^{(l)} (\alpha')$, $\alpha' \in U_{1, 3}$, 
$l \in \mathbb{Z}_{\geqslant 0}$.  
We know that $\Phi_{1}^{(0)} (y_1, x_2, x_3) = S_{1} (y_1, x_2, x_3)$ and 
$\Phi_{1, 3}^{(0)} (y_1, x_2, y_3) = S_{1, 3} (y_1, x_2, y_3)$, 
so the link for the leading coefficients is the Legendre transform. 
What is the link between the higher order coefficients corresponding to $l \geqslant 1$?  

It is more convenient to describe the link mentioned in an abstract setting. 
Consider a Lagrangian manifold $\Lambda^{+} \subset \mathbb{R}^{2 (d + 1)} (x, y)$ 
for an abstract phenomenological thermodynamic system as above. 
We assume that in every focal chart $(U, I)$ of type $I \subset [d + 1]$, 
we are given a function $S_{U, I}^{(\lambda)} (\alpha)$, $\alpha \in U$, depending on a parameter $\lambda > 0$, 
such that there is an asymptotic expansion 
\begin{equation} 
\label{eq:S_U_I_lambda}
S_{U, I}^{(\lambda)} (\alpha) \simeq \sum_{l = 0}^{\infty} \lambda^{- l} \varphi_{U, I}^{(l)} (\alpha), 
\end{equation} 
as $\lambda \to \infty$, where $\varphi_{U, I}^{(l)} (\alpha)$ are smooth functions, $l \in \mathbb{Z}_{\geqslant 0}$. 
Using the notation \eqref{eq:notation_x_y_I}, 
define also the functions $\Phi_{U, I}^{(l)} ((x, y)_{I})$ via 
$\Phi_{U, I}^{(l)} ((x (\alpha), y (\alpha))_{I}) = \varphi_{U, I}^{(l)} (\alpha)$, $\alpha \in U$, 
for $l = 0, 1, 2, \dots$. 
Take another focal chart $(W, J)$ on $\Lambda^{+}$ such that $W \cap U \not = \emptyset$. 
In the leading order $l = 0$, in the points corresponding to $\alpha \in W \cap U$, 
we have the Legendre transform: 
\begin{equation}
\label{eq:Legendre_U_I_W_J}
\Phi_{W, J}^{(0)} |_{\wh{\pi}_{J} (W \cap U)} = \mathcal{L}_{J, I}^{W \cap U} (\Phi_{U, I}^{(0)} |_{\wh{\pi}_{I} (W \cap U)})
\end{equation}
The formulae for the other coefficients mimic basically 
to the formulae of the stationary phase method. 
We need to assume that the Hessian  
\begin{equation*} 
\det \mathrm{Hess} \Phi_{U, I}^{(0)} ((x, y)_{I}) \not = 0, 
\end{equation*}
for every $(x, y)_{I} \in \wh{\pi}_{I} (U)$, for every focal chart $(U, I)$. 

Let us first introduce some notation for the differential operators associated with this method. 
If we have a function $f ((x, y)_{I}) \in C_{0}^{\infty} (\wh{\pi}_{I} (W \cap U))$, 
then we can look at an integral 
$Q_{W \cap U, J, I}^{(\varepsilon)} [\Phi_{U, I}^{(0)}, f] ((x, y)_{J})$, 
$(x, y)_{J} \in \wh{\pi}_{J} (W \cap U)$ depending on a small parameter $\varepsilon > 0$, 
\begin{multline*} 
Q_{W \cap U, J, I}^{(\varepsilon)} [\Phi_{U, I}^{(0)}, f] ((x, y)_{J}) := 
\frac{1}{(2 \pi \varepsilon)^{|J \backslash I| + |I \backslash J|}}
\int 
\Big( \prod_{j \in J \backslash I} d x_j \Big) 
\Big( \prod_{i \in I \backslash J} d y_i \Big)
f ((x, y)_{I}) 
\times \\ \times 
\exp \Big\lbrace \frac{\mathrm{i}}{\varepsilon} \Big[\Phi_{U, I}^{(0)} ((x, y)_{I}) - 
\sum_{i \in I \backslash J} y_i x_i + 
\sum_{j \in J \backslash I} y_j x_j
\Big] \Big\rbrace, 
\end{multline*}
where $|\cdot|$ denotes the cardinality of a set. 
This integral admits an asymptotic expansion as $\varepsilon \to 0$ 
(the \emph{stationary phase method}, see \cite{Fedoryuk}): 
\begin{multline*} 
Q_{W \cap U, J, I}^{(\varepsilon)} [\Phi_{U, I}^{(0)}, f] ((x, y)_{J}) \simeq 
\exp \Big( \frac{\mathrm{i}}{\varepsilon} \Phi_{W, J}^{(0)} ((x, y)_{J}) \Big)
\times \\ \times 
\exp \Big( \frac{\mathrm{i} \pi}{4} \mu_{J, I}^{W \cap U} \Big)
\sum_{n = 0}^{\infty} (- \mathrm{i} \varepsilon)^{n} 
\wh{\mathcal{V}}_{W \cap U, J, I}^{(n)} [\Phi_{U, I}^{(0)}, f] 
(\tau_{I, J}^{W \cap U} ((x, y)_{J})), 
\end{multline*} 
where the map 
$\tau_{I, J}^{W \cap U}: 
\wh{\pi}_{J} (W \cap U) \to \wh{\pi}_{I} (W \cap U)$ is 
the \emph{glueing map} between the focal charts $(U, I)$ and $(W, J)$ on $\Lambda^{+}$
defined by 
$\tau_{I, J}^{W \cap U} ((x (\alpha), y (\alpha))_J) = (x (\alpha), y (\alpha))_{I}$, 
for $\alpha \in W \cap U$, 
$\mu_{J, I}^{W \cap U} \in \mathbb{Z}/ 4 \mathbb{Z}$ is a constant 
(related to the \emph{Maslov index}), 
and 
$\wh{\mathcal{V}}_{W \cap U, J, I}^{(n)} [\Phi_{U, I}^{(0)}, - ]$ are linear partial differential operators, 
such that,
for each $n$,  
$\wh{\mathcal{V}}_{W \cap U, J, I}^{(n)} [\Phi_{U, I}^{(0)}, f] ((x, y)_{I})$ 
depends only on a \emph{finite} number $M_n$ of partial derivatives of $\Phi_{U, I}^{(0)}$ in 
the point $((x, y)_{I}) \in \wh{\pi}_{I} (W \cap U)$.  
The number $M_n$ grows as $n \to \infty$, and in the leading order $n = 0$ the operator 
$\wh{\mathcal{V}}_{W \cap U, J, I}^{(0)} [\Phi_{U, I}^{(0)}, -]$ is just a multiplication over a smooth function, which 
does not vanish in no point of $\wh{\pi}_{I} (W \cap U)$. 
Fix the choice of $\mu_{J, I}^{W \cap U}$ in such a way 
that this function is \emph{positive} 
(this is always possible).

Observe, that if we take instead of $f ((x, y)_{I})$ a function 
$f^{(\varepsilon)} ((x, y)_{I})$ depending on the small parameter $\varepsilon$, such that there is an asymptotic expansion 
$f^{(\varepsilon)} ((x, y)_{I}) \simeq \sum_{m = 0}^{\infty} (- \mathrm{i} \varepsilon)^{m} f_{m} ((x, y)_{I})$, where  
$f_{m} \in C_{0}^{\infty} (\wh{\pi}_{I} (W \cap U))$, $m = 0, 1, 2, \dots$, then, 
using the linearity of the operators 
$\wh{\mathcal{V}}_{W \cap U, J, I}^{(n)} [\Phi_{U, I}^{(0)}, - ]$, $n = 0, 1, 2, \dots$, 
we obtain: 
\begin{multline} 
\label{eq:stationary_phase_f_epsilon}
\exp \Big( - \frac{\mathrm{i}}{\varepsilon} \Phi_{W, J}^{(0)} ((x, y)_{J}) 
- \frac{\mathrm{i} \pi}{4} \mu_{J, I}^{W \cap U} \Big) 
Q_{W \cap U, J, I}^{(\varepsilon)} [\Phi_{U, I}^{(0)}, f^{(\varepsilon)}] ((x, y)_{J})   
\simeq \\ \simeq  
\sum_{n = 0}^{\infty} (- \mathrm{i} \varepsilon)^n 
\sum_{m = 0}^{n}
\wh{\mathcal{V}}_{W \cap U, J, I}^{(n - m)} [\Phi_{U, I}^{(0)}, f_m] 
(\tau_{I, J}^{W \cap U} ((x, y)_{J})), 
\end{multline}
for $(x, y)_{J} \in \wh{\pi}_{J} (W \cap U)$. 
Now, let us assume that $f^{(\varepsilon)}$ can be written in the form 
$f^{(\varepsilon)} = \exp \lbrace g^{(\varepsilon)} \rbrace$, where 
$g^{(\varepsilon)}$ is a convergent power series 
$g^{(\varepsilon)} = \sum_{m = 0}^{\infty} (- \mathrm{i} \varepsilon)^{m} g_{m}/ m!$, where 
$g_m \in C_{0}^{\infty} (\wh{\pi}_{I} (W \cap U))$, 
$m = 0, 1, 2, \dots$. 
Assume also that the left-hand side of \eqref{eq:stationary_phase_f_epsilon} 
(denote it at this moment $F^{(\varepsilon)} ((x, y)_{J})$) 
can also be represented as 
$F^{(\varepsilon)} ((x, y)_{J}) = \exp \big\lbrace G^{(\varepsilon)} (\tau_{I, J}^{W \cap U} ((x, y)_{J})) \big\rbrace$, where 
$G^{(\varepsilon)} ((x, y)_{I})$ is a convergent power series 
$G^{(\varepsilon)} = \sum_{m = 0}^{\infty} (- \mathrm{i} \varepsilon)^{m} G_{m}/ m!$, 
with $G_{m} \in C_{0}^{\infty} (\wh{\pi}_{I} (W \cap U))$, $m = 0, 1, 2, \dots$ 
(at least this does not contradict a possibility of having an asymptotic expansion 
like on the right-hand side of \eqref{eq:stationary_phase_f_epsilon}). 
The link between the collections of the coefficients $\lbrace G_{m} \rbrace_{m = 0}^{\infty}$ and 
$\lbrace g_m \rbrace_{m = 0}^{\infty}$ defines some operators, that we are going to use 
to describe the link between the asymptotic power series \eqref{eq:S_U_I_lambda} 
corresponding to different $(U, I)$.

Take a pair of smooth functions $a (\xi)$ and $b (\xi)$, $\xi$ varies over $\mathbb{R}$, $a (\xi) = \exp (b (\xi))$. 
Look at the Taylor's expansions $a (\xi) = \sum_{m = 0}^{\infty} a_m \xi^m/ m!$ and $b (\xi) = \sum_{n = 0}^{\infty} b_n \xi^n/ n!$. 
For the leading coefficients, $a_0 = \exp (b_0)$. 
If $m \geqslant 1$, then 
\begin{multline*} 
a_m = \Big\lbrace 
\Big( \frac{\partial}{\partial \xi} \Big)^{m} \exp (b (\xi)) 
\Big\rbrace 
\Big|_{\xi = 0} = 
\exp (b_0) 
\Big\lbrace
\Big( \frac{\partial}{\partial \xi} \Big)^{m} 
\sum_{p = 0}^{\infty} \frac{1}{p!} [b (\xi) - b_0]^{p} 
\Big\rbrace
\Big|_{\xi = 0} 
= \\ = 
e^{b_0}
\sum_{p = 1}^{m} \frac{1}{p!}
\sum_{
\substack{I_{1}, I_{2}, \dots, I_{p} \subset [p],\\ 
I_{1}, I_{2}, \dots, I_{p} \not = \emptyset, \\
|I_1| + |I_2| + \dots + |I_{p}| = m
}} 
b_{|I_1|} b_{|I_2|} \dots b_{|I_p|}. 
\end{multline*}
Therefore, $a_m = A_m (b_0, b_1, b_2, \dots, b_m)$, where 
\begin{equation} 
\label{eq:positive_A_m}
A_m (b_0, b_1, b_2, \dots, b_m) := 
e^{b_0} \sum_{p = 1}^{m} \frac{1}{p!} 
\sum_{
\substack{m_1, m_2, \dots, m_p = 1,\\ 
m_1 + m_2 + \dots + m_p = m}}^{m} 
\frac{m!}{m_1! m_2! \dots m_p!} 
b_{m_1} b_{m_2} \dots b_{m_p}, 
\end{equation}
for every $m = 1, 2, \dots$. 
To invert these formulae, observe that $b_0 = \mathrm{ln} a_0$, $a_0 > 0$. 
For $m \geqslant 1$, we have 
\begin{multline*} 
b_{m} = \Big\lbrace 
\Big( \frac{\partial}{\partial \xi} \Big)^{m} \mathrm{ln} (a (\xi)/ a_0)
\Big\rbrace \Big|_{\xi = 0} = 
\Big\lbrace 
\Big( \frac{\partial}{\partial \xi} \Big)^{m} \sum_{q = 1}^{\infty} \frac{(-1)^{q}}{q} [a (\xi)/ a_0 - 1]^{q}
\Big\rbrace \Big|_{\xi = 0} 
= \\ = 
\sum_{q = 1}^{m} \frac{(-1)^{q}}{q a_{0}^{q}}
\sum_{
\substack{J_{1}, J_{2}, \dots, J_{q} \subset [q],\\ 
J_{1}, J_{2}, \dots, J_{q} \not = \emptyset, \\
|J_1| + |J_2| + \dots + |J_{q}| = m
}} 
a_{|J_1|} a_{|J_2|} \dots a_{|J_q|}. 
\end{multline*}
Therefore $b_m = B_m (a_0, a_1, \dots, a_m)$, where  
\begin{equation} 
\label{eq:positive_B_m}
B_{m} (a_0, a_1, \dots, a_m) := 
\sum_{q = 1}^{m} \frac{(-1)^{q}}{q a_{0}^{q}}
\sum_{
\substack{
m_1, m_2, \dots, m_q = 1,\\   
m_1 + m_2 + \dots + m_q = m
}}^{m}
\frac{m!}{m_1! m_2! \dots m_q!}
a_{m_1} a_{m_2} \dots a_{m_q},  
\end{equation}
for every $m = 1, 2, \dots$. 
Extending naturally the notation as 
\begin{equation} 
\label{eq:A_B_0}
A_{0} (b_0) := \exp (b_0), \quad 
B_0 (a_0) := \mathrm{ln} (a_0), 
\end{equation}
we arrive at 
$a_n = A_{n} (b_0, b_1, \dots, b_n)$ and 
$b_{n} = B_{n} (a_0, a_1, \dots, a_n)$ for all $n = 0, 1, 2, \dots$. 

We can now describe how the expansions \eqref{eq:S_U_I_lambda} 
corresponding to different focal charts $(U, I)$ on $\Lambda^{+}$ must be linked. 
Take a pair of focal charts $(U, I)$ and $(W, J)$, $W \cap U \not = \emptyset$, and 
look at the expansions for $S_{U, I}^{(\lambda)} (\alpha)$ and $S_{W, J}^{(\lambda)} (\alpha)$ 
in terms of the local coordinates: 
\begin{equation*} 
\bar S_{U, I}^{(\lambda)} ((x, y)_{I}) \simeq 
\sum_{l = 0}^{\infty} \lambda^{- l} \Phi_{U, I}^{(l)} ((x, y)_{I}), \quad 
\bar S_{W, J}^{(\lambda)} ((x, y)_{J}) \simeq 
\sum_{l = 0}^{\infty} \lambda^{- l} \Phi_{W, J}^{(l)} ((x, y)_{J}), 
\end{equation*}
where 
$(x, y)$ corresponds to a point $\alpha$ varying over $W \cap U$, and 
$\bar S_{U, I}^{(\lambda)} ((x, y)_{I}) := S_{U, I}^{(\lambda)} (\wh{\pi}_{I}^{-1} ((x, y)_{I}))$ 
for $(x, y)_{I} \in \wh{\pi}_{I} (U)$, and 
$\bar S_{W, J}^{(\lambda)} ((x, y)_{J}) := S_{W, J}^{(\lambda)} (\wh{\pi}_{J}^{-1} ((x, y)_{J}))$, 
for $(x, y)_{J} \in \wh{\pi}_{J} (W)$. 
To compute the collection of coefficients $\lbrace \Phi_{W, J}^{(l)} ((x, y)_{J}) \rbrace_{l = 0}^{\infty}$, 
$(x, y)_{J} \in \wh{\pi}_{J} (W \cap U)$, 
from a collection of coefficients $\lbrace \Phi_{U, I}^{(l)} ((x, y)_{I}) \rbrace_{l = 0}^{\infty}$, 
one needs to do the following: 

\begin{itemize} 

\item 
Compute $\Phi_{W, J}^{(0)} ((x, y)_{J})$ as a Legendre transform \eqref{eq:Legendre_U_I_W_J} of $\Phi_{U, I}^{(0)} ((x, y)_{I})$. 

\item 
Set $b_m = \Phi_{U, I}^{(m + 1)} ((x, y)_{I})$, $m = 0, 1, 2, \dots$, and compute the corresponding 
$a_n = A_n (b_0, b_1, \dots, b_n)$, $n = 0, 1, 2, \dots$, 
where $A_n$ is defined in \eqref{eq:positive_A_m} and \eqref{eq:A_B_0}. 

\item 
Compute 
the coefficients of the asymptotic expansion on the right-hand side of 
\eqref{eq:stationary_phase_f_epsilon}, replacing 
$\lbrace f_{m} \rbrace_{m = 0}^{\infty}$ with $\lbrace a_{m} \rbrace_{m = 0}^{\infty}$,  
$a_m = a_m ((x, y)_{I})$: 
\begin{equation*} 
a_{n}' ((x, y)_{J}) := 
\sum_{m = 0}^{n}
\wh{\mathcal{V}}_{W \cap U, J, I}^{(n - m)} [\Phi_{U, I}^{(0)}, a_m] 
(\tau_{I, J}^{W \cap U} ((x, y)_{J})), 
\end{equation*}
for $n = 0, 1, 2, \dots$. 

\item 
Compute $b_n' := B_n (a_1', a_2', \dots, a_n')$, $n = 0, 1, 2, \dots$, 
where $B_n$ is defined in \eqref{eq:positive_B_m} and \eqref{eq:A_B_0}. 
This collection of functions $b_{n}' = b_{n}' ((x, y)_{J})$ 
is precisely the required collection of the higher order coefficients, 
\begin{equation*} 
\Phi_{W, J}^{(n + 1)} ((x, y)_{J}) = b_{n}' ((x, y)_{J}), 
\end{equation*}
where $n = 0, 1, 2, \dots$, and $(x, y)_{J}$ corresponds to a point in $W \cap U$. 

\end{itemize}

It can be of interest to assemble the higher order coefficients $\Phi_{U, I}^{(l)} ((x, y)_{I})$, $l = 1, 2, 3, \dots$, 
into a formal power series 
\begin{equation*} 
\Phi_{U, I}^{(\varepsilon)} ((x, y)_{I}) := 
\sum_{m = 0}^{\infty} \varepsilon^{m} \Phi_{U, I}^{(m + 1)} ((x, y)_{I}), 
\end{equation*} 
where $\varepsilon$ is a formal variable, 
$\Phi_{U, I}^{(\varepsilon)} ((x, y)_{I}) \in C^{\infty} (\wh{\pi}_{I} (U))[[\varepsilon]]$. 
Note, that the coefficient $\Phi_{U, I}^{(0)} ((x, y)_{I})$ is \emph{not} involved in this definition. 
If $U$ admits another type of focal coordinates $J \subset [d + 1]$, then 
the described link between the coefficients in different focal charts yields a \emph{nonlinear} map 
\begin{equation*} 
\mathcal{N}_{J, I}^{U}: C^{\infty} (\wh{\pi}_{I} (U))[[\varepsilon]] \to 
C^{\infty} (\wh{\pi}_{J} (U))[[\varepsilon]]. 
\end{equation*}
We have $\mathcal{N}_{I, I}^{U} = \mathit{id}$, and, in case $U$ admits the focal coordinates of types $I$, $J$, and $K$, 
one can check the cocyclicity condition: 
\begin{equation} 
\label{eq:cocycle_N}
\mathcal{N}_{I, K}^{U} \circ \mathcal{N}_{K, J}^{U} \circ \mathcal{N}_{J, I}^{U} = \mathit{id}. 
\end{equation}
Note, that about $U$ we assume that the Hess's matrices $\mathrm{Hess} S_{I}$ and $\mathrm{Hess} S_{J}$ 
are non-degenerate. 
Intuitively, the map $\mathcal{N}_{J, I}^{U}$ ``extends'' the Legendre transform $\mathcal{L}_{J, I}^{U}$, 
and the cocycle condition \eqref{eq:cocycle_N} corresponds to the cocycle condition \eqref{eq:cocycle_L}. 

Where do the functions $S_{U, I}^{(\lambda)} (\alpha)$ come from? 
Consider again a phenomenological thermodynamic system with 
extensive coordinates $(x_1, x_2, x_3) = (E, V, \nu)$ (internal energy, volume, number of moles), and 
intensive coordinates $(y_1, y_2, y_3) = (\beta, \beta p, - \beta \mu)$ ($\beta$ is the inverse absolute temperature, 
$p$ is pressure, $\mu$ is chemical potential).  
If we know the entropy of the system as a function of extensive coordinates, $S = S (E, V, \nu)$, 
then the first law of thermodynamics can be expressed as follows: 
\begin{equation}
\label{eq:first_law_td}
 d S = \beta d E + (\beta p) d  V + (- \beta \mu) d \nu = \sum_{i = 1}^{3} y_i d x_i, 
\end{equation}
on the Lagrangian manifold $\Lambda^{+}$. 
If $I = \lbrace 1 \rbrace$ and $U \subset \Lambda^{+}$ is fixed, 
then the limit $S_{1} (\alpha) := \lim_{\lambda \to \infty} S_{U, I}^{(\lambda)} (\alpha)$ 
corresponds to the Legendre transform of $S$ in the first argument, 
$d S_1 = (- E d \beta + (\beta p) d V + (- \beta \mu) d \nu)|_{U}$, 
i.e. $S_1 = - \beta F$, where $F$ is the phenomenological free energy described as a function on $U \subset \Lambda^{+}$. 
If one takes the canonical Gibbs distribution 
\begin{equation} 
\label{eq:canonical_Gibbs}
\wh{w}_{k_B, N}^{(L)} (\beta; g) := \frac{1}{\mathcal{Z}_{k_B, N}^{(L)} (\beta; g)} 
\exp \Big( - \frac{\beta \wh{H}_{N}^{(L)} (g)}{k_B} \Big), 
\end{equation}
then 
one can see that the averages $\langle \wh{H}_{N}^{(L)} \rangle$ and 
$\langle ( \wh{H}_{N}^{(L)} - \langle \wh{H}_{N}^{(L)} \rangle)^{2} \rangle$ over 
$\wh{w}_{k_B, N}^{(L)} (\beta; g)$ 
correspond to the derivatives $(- k_B \partial / \partial \beta)^{1}$ and 
$(- k_B \partial / \partial \beta)^{2}$ of $\mathrm{ln} \mathcal{Z}_{k_B, N}^{(L)} (\beta; g)$. 
Substituting the values 
$\beta = \beta_{\lambda} (\alpha)$, 
$L = (V_{\lambda} (\alpha))^{1/ n}$, 
$N = N_{\lambda} (\alpha) = \nu_{\lambda} (\alpha) R/ k_B$, and $g = g_{\lambda} (\alpha)$, 
corresponding to the rescaled system (the rescaling coefficient $\lambda$), 
one extends this fact as follows: the quantities 
\begin{equation*} 
C_{n}^{(\lambda)} (\alpha) := \Big( - k_B \frac{\partial}{\partial \beta} \Big)^{n} 
\frac{\lambda S_{1}^{(\lambda)} (\wh{\pi}_{1}^{-1} (\beta, V, \nu))}{k_B} 
\bigg|_{(\beta, V, \nu) = (\beta_{\lambda} (\alpha), V_{\lambda} (\alpha), \nu_{\lambda} (\alpha))}
\end{equation*}
define the \emph{cumulants} of the fluctuations of energy, $n = 1, 2, 3, \dots$, 
where $\wh{\pi}_{1} : \alpha \mapsto (\beta (\alpha), V (\alpha), \nu (\alpha))$ is the canonical projection. 
Observe, that $C_1^{(\lambda)} (\alpha) = O (\lambda)$, $\lambda \to \infty$, which is the \emph{expectation value} of internal energy, 
and $C_2^{(\lambda)} (\alpha) = O (\lambda)$, the \emph{variance} of the corresponding fluctuations, 
so the standard deviation is $O (\sqrt{\lambda})$, as it should be in quasithermodynamics. 
The inverse absolute temperature $\beta$ is \emph{fixed} by the thermostat, and the canonically conjugate \emph{extensive} 
quantity (the internal energy) \emph{fluctuates}. 
This interpretation is naturally dualized: 
if an extensive quantity is fixed (for example, the volume $V$), 
then the canonically conjugate intensive quantity fluctuates. 
From the first law of thermodynamics \eqref{eq:first_law_td}, we can see, for example, that the 
derivatives over $V$ should define the cumulants $\wt{C}_{m}^{(\lambda)} (\alpha)$ of the fluctuations of $\beta p$, 
where $p$ is the pressure in the system. 
More precisely, the derivative $\partial / \partial \beta$ should be replaced with $\lambda^{-1} \partial/ \partial V$, 
since $V_{\lambda} (\alpha) = O (\lambda)$ (extensive variable), and 
$\beta_{\lambda} (\alpha) = O (1)$ (intensive variable), $\lambda \to \infty$, 
and we should take into account the minus sign corresponding to the Legendre transform: 
\begin{equation*} 
\wt{C}_{m}^{(\lambda)} (\alpha) = \Big( k_B \lambda^{-1} \frac{\partial}{\partial V} \Big)^{m} 
\frac{\lambda S_{1}^{(\lambda)} (\wh{\pi}_{1}^{-1} (\beta, V, \nu)))}{k_B}
\bigg|_{(\beta, V, \nu) = (\beta_{\lambda} (\alpha), V_{\lambda} (\alpha), \nu_{\lambda} (\alpha))}, 
\end{equation*} 
for $m = 1, 2, 3, \dots$. 
In particular, the first two cumulants 
for the fluctuations of $\beta p$ 
(the expectation value and the variance) satisfy  
$C_{1}^{(\lambda)} (\alpha) = O (1)$ and $C_{2}^{(\lambda)} (\alpha) = O (\lambda^{-1})$, $\lambda \to \infty$, 
so the standard deviation corresponding to $\beta p$ is $O (1/ \sqrt{\lambda})$, as it should be in quasithermodynamics.

For an abstract phenomenological thermodynamic system 
$\Lambda^{+} \subset \mathbb{R}^{2 (d + 1)} (x, y)$ the interpretation 
of the functions $S_{U, I}^{(\lambda)} (\alpha)$ for every focal chart $(U, I)$ is as follows. 
We know that $[d + 1] \backslash I$ cannot in thermodynamics be empty (by construction). 

\begin{itemize} 

\item
The first law of thermodynamics is expressed over $(U, I)$ as follows:  
\begin{equation*} 
d \varphi_{U, I}^{(0)} = 
\Big( - \sum_{i \in I} x_i d y_i + \sum_{j \in [d + 1] \backslash I} y_j d x_j \Big) \Big|_{U}, 
\end{equation*}
where 
$\varphi_{U, I}^{(0)} (\alpha) := \lim_{\lambda \to \infty} S_{U, I}^{(\lambda)} (\alpha)$, $\alpha \in U \subset \Lambda^{+}$, 
$x = (x_1, x_2, \dots, x_{d + 1})$ are the extensive coordinates, and 
$y = (y_1, y_2, \dots, y_{d + 1})$ are the canonically conjugate intensive coordinates. 
A model example is $d = 2$,  
$(x_1, x_2, x_3) = (E, V, \nu)$ (internal energy, volume, number of moles), 
$(y_1, y_2, y_3) = (\beta, \beta p, - \beta \mu)$ (inverse absolute temperature $\beta$, pressure $p$, 
chemical potential $\mu$)

\item 
The units of measurement of $S_{U, I}^{(\lambda)} (\alpha)$ are same as the units of measurement of 
the Boltzmann constant $k_B$, $[S_{U, I}^{(\lambda)} (\alpha)] = [k_B]$. 
The function $\varphi_{U, \emptyset}^{(0)} (\alpha)$ corresponding to the case $I = \emptyset$ 
is the \emph{entropy} of the phenomenological system in the state $\alpha \in U \subset \Lambda^{+}$.

\item 
The following collection of derivatives 
is interpreted as the \emph{cumulants} 
describing the fluctuations of 
the vector $(x, y)_{[d + 1] \backslash I}$
over a state $\alpha \in U$: 
\begin{multline*} 
C_{M}^{(U, I, \lambda)} (\alpha) 
\big|_{\alpha = \wh{\pi}_{U, I}^{-1} ((x, y)_{I})} := \\
\Big( \prod_{i \in I} \Big(- k_B \frac{\partial}{\partial y_i} \Big)^{m_i} \Big)
\Big( \prod_{j \in [d + 1] \backslash I} \Big( k_B \lambda^{-1} \frac{\partial}{\partial x_j} \Big)^{m_j} \Big)
\frac{\lambda S_{U, I}^{(\lambda)} (\wh{\pi}_{U, I}^{-1} (x, y)_{I})}{k_B}, 
\end{multline*}
where 
$M = (m_1, m_2, \dots, m_{d + 1}) \in \mathbb{Z}_{\geqslant 0}^{d + 1}$ is an integer multi-index, 
$M \not = (0, 0, \dots, 0)$, 
and $\wh{\pi}_{U, I}: U \ni \alpha \mapsto (x (\alpha), y (\alpha))_{I}$ denotes the canonical projection, 
$(x, y)_{I} \in \wh{\pi}_{I} (U)$. 
The cumulants corresponding to the same $M \in \mathbb{Z}_{\geqslant 0}^{d + 1}$, but 
computed over two different focal charts $(U, I)$and $(W, J)$, 
(i.e. \emph{different} families of rescaled systems)
coincide asymptotically in the intersection: 
\begin{equation*} 
C_{M}^{(U, I, \lambda)} (\alpha) = 
C_{M}^{(W, J, \lambda)} (\alpha) + O (\lambda^{- \infty}), 
\end{equation*}
as $\lambda \to \infty$, 
where $\alpha \in U \cap W$. 

\end{itemize}

The cocycle conditions \eqref{eq:cocycle_L} basically say that one can glue a 
Lagrangian manifold $\Lambda^{+}$. Informally speaking, \emph{phenomenological thermodynamics} IS 
a Lagrangian manifold $\Lambda^{+}$ (connected, simply connected and admitting projectivization 
with respect to extensive coordinates). 
The \emph{statistical thermodynamics} IS the Gibbs distribution
corresponding to a focal chart $(U, I)$ on $\Lambda^{+}$, $I \subset [d + 1]$, $I \not = [d + 1]$, 
for example, the canonical distribution \eqref{eq:canonical_Gibbs} if $I = \lbrace i_0 \rbrace$ 
is a singleton and $i_0$ corresponds to the inverse absolute temperature. 
The type $I$ of the focal chart describes the way we ``extract'' the system from the outside world 
(fixed adiabatic walls, fixed heat-conducting walls, a non-fixed piston, walls penetrable to certain sorts of particles, etc.)  
This way of extraction does matter if we consider the Gibbs distribution along the families of rescaled systems 
by a parameter $\lambda \to \infty$. 
If one is ready to sacrifice the precision modulo $O (\lambda^{- \infty})$ describing the 
fluctuations of the intensive and extensive quantities, then one arrives 
at ``quasithermodynamics''. 
One may say that \emph{quasithermodynamics} IS a global section of the sheaf 
on $\Lambda^{+}$ corresponding to the cocycle conditions \eqref{eq:cocycle_N} 
(the formal variable $\varepsilon$ corresponds to the small parameter $\lambda^{-1}$, 
but should not be confused with it). 
This sheaf corresponds to the sheaf of $\mathcal{V}$-objects in \cite{Maslov_op_meth}, and the 
analogy with mechanics is as follows  \cite{Maslov_thermo1}: 
phenomenological thermodynamics corresponds to \emph{classical} mechanics, 
statistical thermodynamics corresponds to \emph{quantum} mechanics, and 
quasithermodynamics corresponds to \emph{semiclassical} mechanics.

\section{Nonequilibrium quasithermodynamics} 

We are now interested in extending the quasithermodynamics picture to a nonequilibrium setting. 
One may use the concept of \emph{relevant ensembles} for this purpose \cite{ZubarevMorozovRopke_vol1}. 
Like in the previous section, to keep the story more simple, 
consider a multi-particle quantum mechanical system on a $n$-dimensional torus of radius $L$. 
Denote $\mathcal{F}$ the Fock space of the system, and $\wh{\mathcal{H}}$ the Hamiltonian of the system 
(a self-adjoint operator on $\mathcal{F}$). 
Fix a collection of observables 
$\wh{\mathcal{E}}_{1}^{(L)} (g), \wh{\mathcal{E}}_{2}^{(L)} (g), \dots, \wh{\mathcal{E}}_{d}^{(L)} (g)$ 
(self-adjoint operators on the Fock space of the system), depending on a 
collection of parameters $g = (g_1, g_2, \dots, g_m)$. 
One considers the Gibbs's statistical operator 
\begin{equation} 
\label{eq:Gibbs_operator}
\wh{w}_{k_B}^{(L)} (\beta; g) := \frac{1}{\mathcal{Z}_{k_B}^{(L)} (\beta; g)} 
\exp \Big( - \frac{1}{k_B} \sum_{i = 1}^{d} \beta_{i} \wh{\mathcal{E}}_{i}^{(L)} (g) \Big), 
\end{equation}
where $\mathcal{Z}_{k_B}^{(L)} (\beta; g) :=  
\mathrm{Tr} \big( - \sum_{i = 1}^{d} \beta_i \wh{\mathcal{E}}_{i}^{(L)} (g)/ k_B \big)$ 
(we assume that the trace exists), 
$k_B$ is the Boltzmann constant, and $\beta = (\beta_1, \beta_2, \dots, \beta_d)$. 
Set 
\begin{equation*} 
\Phi_{k_B}^{(L)} (\beta; g) := k_B \mathrm{ln} \mathcal{Z}_{k_B}^{(L)} (\beta; g), 
\end{equation*}
The general idea is that the evolution of the system is 
well approximated by a family of operators 
$\big\lbrace \wh{w}_{k_B}^{(L^{t})} (\beta^{t}, g^{t}) \big\rbrace_{t}$, 
where $t$ is time, and $L^t$, $\beta^t$, $g^t$ are certain functions 
(we let $L$ and $g$ change with time for more generality). 
One says that the state $\mathrm{Tr} \big\lbrace \wh{w}_{k_B}^{(L^t)} (\beta^t; g^t) - \big\rbrace$ 
defines a \emph{relevant ensemble} at the moment of time $t$.

Let us look at the collection of all possible relevant ensembles, i.e. all possible 
statistical operators $\wh{w}_{k_B}^{(L)} (\beta; g)$. 
To mimic the setup of the equilibrium thermodynamics, 
consider a phase space $\mathbb{R}_{x, y}^{2 (d + 1)}$ with symplectic structure 
$\omega^{+} := \sum_{i = 1}^{d + 1} d y_i \wedge d x_i$, $x = (x_1, x_2, \dots, x_{d + 1})$, 
$y = (y_1, y_2, \dots, y_{d + 1})$. 
The first $d$ coordinates $x_i$ correspond to $\wh{\mathcal{E}}_{i}^{(L)} (g)$, $i \leqslant d$, and 
the last coordinate $x_{d + 1}$ corresponds to the volume $V = L^{n}$. 
The first $d$ coordinates $y_i$ correspond to $\beta_i$, $i \leqslant d$, 
and the last coordinate $y_{d + 1}$ can be perceived as a non-equilibrium analogue of the 
``pressure over temperature'' variable in the equilibrium thermodynamics. 
The coordinates $x$ are termed the \emph{extensive} coordinates, and 
the coordinates $y$ are termed the \emph{intensive} coordinates. 
Assume that there exists a Lagrangian manifold $\Lambda^{+} \subset \mathbb{R}_{x, y}^{2 (d + 1)}$, 
which is connected, simply connected, 
admits projectivization with respect to the 
extensive coordinates (i.e. the analogue of the Lagrangian manifold in phenomenological equilibrium thermodynamics). 
Assume, for simplicity, that $\Lambda^{+}$ admits $(y_1, y_2, \dots, y_d, x_{d + 1})$ 
as global coordinates (i.e. is covered by a single focal chart of type $[d] \subset [d + 1]$). 
Let $\lambda \to \infty$ be a large parameter (termed the \emph{rescaling parameter}), and 
assume that there exist functions $C (\lambda)$ and 
$\beta_{\lambda} (\alpha)$, $g_{\lambda} (\alpha)$, where $\alpha$ varies over $\Lambda^{+}$, 
such that 
\begin{equation} 
\label{eq:Phi_C_asymptotic}
\lambda^{-1} \Phi_{k_B}^{((\lambda x_{d + 1} (\alpha))^{1/ n})} (\beta_{\lambda} (\alpha); g_{\lambda} (\alpha)) - 
C (\lambda) 
\simeq 
\sum_{l = 0}^{\infty}
\lambda^{- l} \phi_{k_B}^{(l)} (\alpha), 
\end{equation}
where 
$\phi_{k_B}^{(l)} (\alpha)$ are smooth functions, $l = 0, 1, 2, \dots$. 
This expansion should be perceived in analogy with \eqref{eq:S_U_I_lambda}, 
and $V_{\lambda} (\alpha) = \lambda x_{d + 1} (\alpha)$ 
corresponds to the expanding volume as $\lambda \to \infty$.

Like in the equilibrium case, we should draw a distinction between 
\emph{phenomenological}, \emph{statistical}, and ``\emph{quasi}'' nonequilibrium thermodynamics. 
The general idea is that the phenomenological nonequilibrium thermodynamics 
is described as a curve $\gamma = \lbrace \alpha^{t} \rbrace_{t}$ on the Lagrangian manifold $\Lambda^{+}$, 
where $t$ is time. If this is a curve of relaxation to equilibrium, then 
some points of $\Lambda^{+}$ should correspond to the equilibrium states. 
As a model example, one may think of a system consisting of two subsystems 
which interchange heat with each other, 
but have different temperatures
(the volumes of the subsystems are assumed to be fixed). 
The internal energies $E_1$ and $E_2$ of these two subsystems 
(indexed by 1 and 2) will change 
in such a way that $E_1 + E_2$ is fixed, 
until the inverse temperatures $\beta_1$ and $\beta_2$ become equal. 
More generally, if, for example, the total volume of the system $V (\alpha) = x_{d + 1} (\alpha)$ 
is fixed along the evolution $\gamma = \lbrace \alpha^{t} \rbrace_{t}$, 
$V (\alpha^t) = \mathrm{const}$, 
then under some assumptions, like ``the smallness of time memory effects'' \cite{ZubarevMorozovRopke_vol1}, one 
can derive the following system of equations for $E_{i} (t) := x_{i} (\alpha^t)$, $i = 1, 2, \dots, d$: 
\begin{equation} 
\label{eq:Onsager}
\frac{\partial E_i (t)}{\partial t} = J_i (\alpha^t) + \sum_{j = 1}^{n} L_{i, j} (\alpha^t) \beta_{j} (t), 
\end{equation}
where 
$\beta_i (t) = y_i (\alpha^t)$, $i \in [d]$, 
$L_{i, j} (\alpha)$ are the \emph{Onsager coefficients} corresponding to $\alpha \in \Lambda^{+}$ 
(in practice, they are often approximated as constants)
, and 
$J_{i} (\alpha^t)$, $i \in [d]$, are the \emph{flows} 
induced by $\lbrace \beta_j (t) \rbrace_{j \in [d]}$. 
These flows can be perceived, for each $i \in [d]$, as a limit 
\begin{equation} 
\label{eq:flows_phenomen}
J_{i} (\alpha) := \lim_{\lambda \to \infty} 
\Big( \lambda^{-1} 
\mathrm{Tr} \big\lbrace \wh{w}_{k_B}^{(L)} (\beta; g) 
\wh{\mathcal{J}}_{i}^{(L)} (g)
\big\rbrace 
\Big|_{L = (\lambda x_{d + 1} (\alpha))^{1/ n}, \beta = \beta_{\lambda} (\alpha), g = g_{\lambda} (\alpha)}
\Big), 
\end{equation}
where $\wh{\mathcal{J}}_{i}^{(L)} (g)$ are the \emph{operators of flows}, 
\begin{equation} 
\label{eq:flow_operators}
\wh{\mathcal{J}}_{i}^{(L)} (g) := \frac{\mathrm{i}}{\hbar} \Big[ 
\wh{\mathcal{H}}^{(L)} (g), \wh{\mathcal{E}}_{i}^{(L)} (g)
\Big], 
\end{equation}
the square bracket denotes a commutator, and $\hbar$ is the Planck's constant. 

Denote the left-hand side of \eqref{eq:Phi_C_asymptotic} as $S_{k_B}^{(\lambda)} (\alpha)$, 
and let $\bar S_{k_B}^{(\lambda)} (y_1, y_2, \dots, y_d, x_{d + 1})$ be this 
function expressed in terms of the focal coordinates on $\Lambda^{+}$ of type $[d] \subset [d + 1]$, 
\begin{equation*} 
\bar S_{k_B}^{(\lambda)} (y_1 (\alpha), y_2 (\alpha), \dots, y_d (\alpha), x_{d + 1} (\alpha)) = 
S_{k_B}^{(\lambda)} (\alpha), 
\end{equation*}
for $\alpha \in \Lambda^{+}$. 
In analogy with the previous section, one interprets the derivatives 
\begin{equation} 
\label{eq:cumulants_global}
C_{M}^{(\lambda)} (\alpha) := 
\bigg(
\Big( \prod_{i = 1}^{d} \Big( - k_B \frac{\partial}{\partial y_i} \Big)^{m_i} \Big) 
\Big( k_B \lambda^{-1} \frac{\partial}{\partial x_{d + 1}} \Big)^{m_{d + 1}} 
\frac{\lambda \bar S_{k_B}^{(\lambda)} (y_1, \dots, y_d, x_{d + 1})}{k_B}    \bigg) \bigg|_{\Lambda^{+}} (\alpha), 
\end{equation}
where $\alpha \in \Lambda^{+} \subset \mathbb{R}_{x, y}^{2 (d + 1)}$, 
$M = (m_1, \dots, m_d, m_{d + 1}) \in \mathbb{Z}_{\geqslant 0}^{d + 1}$, 
$M \not = (0, 0, \dots, 0)$, 
as the \emph{cumulants} of fluctuations of 
$(x_1, \dots, x_d, y_{d + 1})$. 
The Lagrangian manifold $\Lambda^{+}$ is recovered from 
the leading term $\bar \Phi_{k_B}^{(0)} (y_1, \dots, y_d, x_{d + 1}) := 
\lim_{\lambda \to \infty} \bar S_{k_B}^{(\lambda)} (y_1, \dots, y_d, x_{d + 1})$ 
of the asymptotic expansion in $\lambda^{-1}$ as a system of equations: 
\begin{equation*} 
x_{i} = - \frac{\partial \bar \Phi_{k_B}^{(0)} (y_1, \dots, y_d, x_{d + 1})}{\partial y_i}, \quad 
y_{d + 1} = \frac{1}{x_{d + 1}} \bar \Phi_{k_B}^{(0)} (y_1, \dots, y_d, x_{d + 1}), 
\end{equation*}
where $i = 1, 2, \dots, d$. 

It follows that in every point $\alpha^t$ of the curve of the evolution $\gamma = \lbrace \alpha^t \rbrace_t$, 
we have a collection of cumulants $\lbrace C_{M}^{(\lambda)} (\alpha^t) \rbrace_{M}$ 
describing the fluctuations of the measured values of 
the quantities associated with $\wh{\mathcal{E}}_{1}^{(L)} (g), \wh{\mathcal{E}}_{2}^{(L)} (g), \dots, \wh{\mathcal{E}}_{d}^{(L)} (g)$ and  $\wt{p}$, 
where $\wt{p}$ is the intensive quantity dual to the volume
(the nonequilibrium ``pressure over temperature''). 
This is, of course, an approximation, depending, essentially, on a 
successful choice of the basis variables $\wh{\mathcal{E}}_{i}^{(L)} (g)$, $i \in [d]$, for the family of relevant ensembles.  
These collections of cumulants stem from the \emph{same} Lagrangian manifold $\Lambda^{+}$, but, in fact, 
one can be more general, and consider \emph{germs} of different Lagrangian manifolds 
``attached'' to each point $\alpha^t \in \gamma \subset \Lambda^{+} \subset \mathbb{R}_{x, y}^{2 (d + 1)}$.  
To see this, one needs to consider the \emph{generalized} Fokker-Planck equation \cite{ZubarevMorozovRopke_vol2}. 
The idea of the derivation of this equation is totally similar to the one for \eqref{eq:Onsager}, 
except that 
instead of a \emph{finite} collection of basis quantities 
$\lbrace \wh{\mathcal{E}}_{i}^{(L)} (g) \rbrace_{i \in [d]}$, one 
considers a \emph{continuous} collection 
$\lbrace L^{n} \delta_{\varepsilon} (a_1 - \wh{\mathcal{E}}_{1}^{(L)} (g)) \dots 
\delta_{\varepsilon} (a_d - \wh{\mathcal{E}}_{d}^{(L)} (g)) \rbrace_{(a_1, \dots a_d) \in \mathbb{R}^{d}}$, 
where $\delta_{\varepsilon} (\cdot)$ is a smooth approximation of the Dirac's delta, 
$\delta_{\varepsilon} (\cdot) \to \delta (\cdot)$ as $\varepsilon \to 0$ (in the weak sense), 
$\int d a \, \delta_{\varepsilon} (a) = 1$, 
$a = (a_1, a_2, \dots, a_d)$. 
Of course, a special care needs to be taken about treating a delta function of an operator, and 
also there is a technical problem related to the noncommutativity of 
the operators 
$\wh{\mathcal{E}}_{i}$, $i \in [d]$. 
Let us assume, for simplicity, that $\wh{\mathcal{E}}_{1}^{(L)} (g), \wh{\mathcal{E}}_{2}^{(L)} (g), \dots, 
\wh{\mathcal{E}}_{d}^{(L)} (g)$ 
mutually commute 
(the noncommutativity can be handled as well, see, for example, \cite{Morozov}). 
Then, for every  $a = (a_1, a_2, \dots, a_d)$, we can just write 
\begin{equation*} 
\wh{\mathcal{F}}_{\varepsilon}^{(L)} (a; g) := L^{n} \prod_{i = 1}^{d} 
\delta_{\varepsilon} (a_i - \wh{\mathcal{E}}_{i}^{(L)} (g)), 
\end{equation*}
without paying attention to the order of operators. 
Assume 
that the corresponding operators of flows 
$\wh{\mathcal{J}}_{1}^{(L)} (g), \wh{\mathcal{J}}_{2}^{(L)} (g), \dots, \wh{\mathcal{J}}_{d}^{(L)} (g)$
defined in \eqref{eq:flow_operators} mutually commute as well. 

In place of $\lbrace \wh{\mathcal{E}}_{i}^{(L)} (g) \rbrace_{i \in [d]}$, we have a collection 
$\lbrace \wh{\mathcal{F}}_{\varepsilon}^{(L)} (a; g) \rbrace_{a \in \mathbb{R}^{d}}$, but 
with an additional condition: 
\begin{equation*} 
\int d a \, \wh{\mathcal{F}}_{\varepsilon}^{(L)} (a; g) = L^{n}.  
\end{equation*}
One must also point out, that all $\mathcal{F}_{\varepsilon}^{(L)} (a; g)$, $a \in \mathbb{R}^{d}$, 
have the same units of measurement. 
In this sense, $\lbrace \wh{\mathcal{F}}_{\varepsilon}^{(L)} (a; g) \rbrace_{a \in \mathbb{R}^{d}}$ 
corresponds to $\lbrace \wh{\mathcal{E}}_{i}^{(L)} (g) \rbrace_{i = 0}^{d}$, where 
$\wh{\mathcal{E}}_{0}^{(L)} (g) := L^n - \sum_{i = 1}^{d} \wh{\mathcal{E}}_{i}^{(L)} (g)$, 
if we choose all $\wh{\mathcal{E}}_{i}^{(L)} (g)$, $i \in [d]$,  
having the same units of measurement as the volume $L^n$, so that we have a right to add them. 
The Gibbs's statistical operator \eqref{eq:Gibbs_operator} can be expressed as follows: 
\begin{equation*} 
\wh{w}_{k_B}^{(L)} (\beta; g) = 
\exp \bigg\lbrace
- \frac{1}{k_B} \bigg[ 
\frac{\Phi_{k_B}^{(L)} (\beta; g)}{L^n} \wh{\mathcal{E}}_{0}^{(L)} (g) + 
\sum_{i = 1}^{d} \Big( 
\beta_i + \frac{\Phi_{k_B}^{(L)} (\beta; g)}{L^n}
\Big) \wh{\mathcal{E}}_{i}^{(L)} (g)
\bigg]
\bigg\rbrace. 
\end{equation*}
Denote  
\begin{equation} 
\label{eq:Gibbs_eta}
\wh{W}_{k_B}^{(L)} (\eta; g) := 
\exp \Big( 
- \frac{1}{k_B} \sum_{i = 0}^{d} \eta_i \wh{\mathcal{E}}_{i}^{(L)} (g)
\Big), 
\end{equation}
where $\eta = (\eta_0, \eta_1, \dots, \eta_d)$ is a collection of parameters. 
Observe that the trace of this operator is equal to one if we substitute 
$\eta_0 = L^{-n} \Phi_{k_B}^{(L)} (\beta; g)$, and $\eta_i = \beta_i + L^{-n} \Phi_{k_B}^{(L)} (\beta; g)$, $i \in [d]$.  
One can also check, that 
\begin{equation*} 
\Big( \sum_{i = 0}^{d} \frac{\partial}{\partial \eta_i} \Big) 
\mathrm{ln} \, \mathrm{Tr} \big( \wh{W}_{k_B}^{(L)} (\eta; g) \big) = L^{n}, 
\end{equation*}
for any $\eta = (\eta_0, \eta_1, \dots, \eta_d)$. 
Consider a phase space with coordinates $\mathbb{R}_{\xi, \eta}^{2 (d + 1)}$ with coordinates 
$\eta = (\eta_0, \eta_1, \dots, \eta_d)$, and $\xi = (\xi_0, \xi_1, \dots, \xi_d)$, 
and symplectic structure $\bar \omega := \sum_{i = 0}^{d} d \eta_i \wedge d \xi_i$. 
The map $\tau: \mathbb{R}_{\xi, \eta}^{2 (d + 1)} \to \mathbb{R}_{x, y}^{2 (d + 1)}$, 
$(\xi, \eta) \mapsto (x, y)$, 
defined by 
$x_i = \xi_i$, $y_i = \eta_i - \eta_{0}$, for $0 \leqslant i \leqslant d$, and 
$x_{d + 1} = \xi_{0} + \sum_{i = 1}^{d} \xi_i$, $y_{d + 1} = \eta_{0}$, is symplectic. 
Therefore, in place of the Lagrangian manifold $\Lambda^{+} \subset \mathbb{R}_{x, y}^{2 (d + 1)}$ 
one can work in terms of the manifold $\tau^{-1} (\Lambda^{+}) \subset \mathbb{R}_{\xi, \eta}^{2 (d + 1)}$, which is 
also Lagrangian. 
If one fixes the collection of parameters $g$ and considers a Lagrangian manifold 
$\Lambda_{g}^{+}$ defined by equations 
\begin{equation*} 
\begin{gathered}
(x_i - \mathrm{Tr} (\wh{\mathcal{E}}_{i}^{(L)} (g) \wh{w}_{k_B}^{(L)} (\beta; g))
|_{\beta = (y_1, \dots, y_d), L= x_{d + 1}^{1/ n}} )|_{\Lambda_{g}^{+}} = 0, \\
\Big( y_{d + 1} - \frac{\partial}{\partial x_{d + 1}} k_B \mathrm{ln} \mathcal{Z}_{k_B}^{(x_{d + 1}^{1/ n})} (\beta; g)
\Big|_{\beta = (y_1, \dots, y_d)} \Big) \Big|_{\Lambda_{g}^{+}} = 0, 
\end{gathered}
\end{equation*}
where $i = 1, 2, \dots, d$ 
(i.e. the generating function is $k_B \mathrm{ln} \mathcal{Z}_{k_B}^{(x_{d + 1}^{1/ n})} ((y_1, \dots, y_d); g)$ 
and the focal coordinates are $(y_1, \dots, y_d, x_{d + 1})$), then for the manifold $\tau^{-1} (\Lambda_{g}^{+})$ 
one can claim that 
\begin{equation*} 
\Big( \xi_{i} + k_B \frac{\partial}{\partial \eta_i} 
\mathrm{Tr} \wh{W}_{k_B}^{(L)} (\eta; g) \Big|_{L = (\sum_{j = 0}^{d} \xi_j)^{1/ n}} \Big) \Big|_{\tau^{-1} (\Lambda_{g}^{+})} = 0, 
\end{equation*}
for $i = 0, 1, \dots, d$. 
These $d + 1$ equations do not characterize 
$\tau^{-1} (\Lambda_{g}^{+})$ completely, since the dimension of the manifold is $d + 1$, 
while the sum of all these equations yields an identity $0 \equiv 0$. 
The missing condition is the normalization: 
\begin{equation*} 
\Big( 
\mathrm{Tr} \wh{W}_{k_B}^{(L)} (\eta; g) \Big|_{L = (\sum_{j = 0}^{d} \xi_j)^{1/ n}} 
\Big) \Big|_{\tau^{-1} (\Lambda_{g}^{+})} = 1. 
\end{equation*}

Now let us look at the continuous case. 
The analogue of the Gibbs's statistical operator \eqref{eq:Gibbs_operator} is as follows: 
\begin{equation*} 
\wh{W}_{k_B, \varepsilon}^{(L)} (\sigma (\cdot); g) := 
\exp \Big( - \frac{1}{k_B} \int_{\mathbb{R}^{d}} d a \, \sigma (a) \wh{\mathcal{F}}_{\varepsilon}^{(L)} (a; g) \Big), 
\end{equation*}
where $\sigma (a)$ is a smooth function that we use instead of $\lbrace \eta_{i} \rbrace_{i = 0}^{d}$ in the formula 
\eqref{eq:Gibbs_eta}. 
An analogue of the phase space $\mathbb{R}_{\xi, \eta}^{2 (d + 1)}$ is formed by pairs of functions 
$(F (\cdot), \sigma (\cdot))$, equipped with a canonical symplectic structure 
$\wt{\omega} := \int d z \, (\delta \sigma) (z) \wedge (\delta F) (z)$. 
An analogue of the manifold $\Lambda_{g}^{+}$, which we denote $\wt{\Lambda}_{\varepsilon, g}^{+}$ is described as follows. 
For every $L > 0$, look at 
\begin{equation*} 
\mathcal{Y}_{\varepsilon}^{(L)} (g) := \lbrace \sigma (\cdot) \,|\, 
\mathrm{Tr} \wh{W}_{k_B, \varepsilon}^{(L)} (\sigma (\cdot); g) = 1
\rbrace. 
\end{equation*}
Then $\wt{\Lambda}_{\varepsilon, g}^{+}$ is described as 
\begin{multline*} 
\wt{\Lambda}_{\varepsilon, g}^{+} = \Big\lbrace 
(F (\cdot), \sigma (\cdot)) \,\Big|\, 
F (a) = \Big(- k_B \frac{\delta}{\delta \sigma (a)} \Big) 
\mathrm{Tr} \wh{W}_{k_B, \varepsilon}^{(L)} (\sigma (\cdot); g), 
a \in \mathbb{R}^{d}, \\ 
\sigma (\cdot) \in \mathcal{Y}_{\varepsilon}^{(L)} (g), L > 0
\Big\rbrace. 
\end{multline*}
Since we are interested in the limit $\varepsilon \to 0$ 
(i.e. the approximation $\delta_{\varepsilon} (\cdot)$ becomes the Dirac's delta), 
one can proceed as 
\begin{multline*} 
\mathrm{Tr} 
\wh{W}_{k_B, \varepsilon}^{(L)} (\sigma (\cdot); g) = 
\mathrm{Tr} 
\bigg\lbrace 
\int d a'\, 
\exp \Big(- \frac{L^n}{k_B} \int d a\, \sigma (a) \delta_{\varepsilon} (a - a') \Big) 
\times \\ \times
\prod_{j = 1}^{d} \delta_{\varepsilon} (a_j' - \wh{\mathcal{E}}_{j}^{(L)} (g)) 
\bigg\rbrace + o (\varepsilon) 
= 
\int d a 
\exp \Big(- \frac{L^n}{k_B} \sigma (a) \Big) 
\Gamma_{\varepsilon}^{(L)} (a; g)
 + o (\varepsilon), 
\end{multline*}
where 
\begin{equation*} 
\Gamma_{\varepsilon}^{(L)} (a; g) := 
\mathrm{Tr} \Big( \prod_{j = 1}^{d} \delta_{\varepsilon} (a_j - \wh{\mathcal{E}}_{j}^{(L)} (g)) \Big)
\end{equation*}
is the analogue of the \emph{statistical weight} of a microcanonical distribution. 
Therefore, the equations describing $\wt{\Lambda}_{\varepsilon, g}^{+}$ in the limit $\varepsilon \to 0$ become more simple. 
Computing the variational derivative, for every $a = (a_1, a_2, \dots, a_d)$, we obtain:  
\begin{equation*} 
F (a) = L^n \exp \Big( - \frac{L^n}{k_B} \sigma (a) \Big) \Gamma_{\varepsilon}^{(L)} (a; g) + o (\varepsilon), 
\end{equation*}
where $\sigma (\cdot)$ varies over $\mathcal{Y}_{\varepsilon}^{(L)} (g)$ as $L$ varies over $\mathbb{R}_{> 0}$.  
This motivates the following construction for the \emph{phenomenological} Lagrangian manifold, 
which we denote just $\wt{\Lambda}^{+}$. 
We assume, that we are given a function $\Gamma^{(L)} (a)$, $a = (a_1, a_2, \dots, a_d)$, depending on a 
parameter $L > 0$. 
Set 
\begin{equation*} 
\mathcal{Y}^{(L)} := \lbrace \sigma (\cdot) \,|\, 
\int d a 
\exp \Big(- \frac{L^n}{k_B} \sigma (a) \Big) 
\Gamma^{(L)} (a) = 1
\rbrace, 
\end{equation*}
for every $L > 0$. 
The Lagrangian manifold $\wt{\Lambda}^{+}$ is described as a collection of pairs, 
\begin{equation} 
\label{eq:Lambda_square}
\wt{\Lambda}^{+} := 
\bigcup_{L > 0}
\Big\lbrace 
(F (\cdot), \sigma (\cdot)) \,|\, 
\sigma (\cdot) \in \mathcal{Y}^{(L)} \,\&\, 
\forall a: 
F (a) =  L^n \exp \Big( - \frac{L^n}{k_B} \sigma (a) \Big) \Gamma^{(L)} (a)
\Big\rbrace. 
\end{equation} 
This manifold is an analogue of $\tau^{-1} (\Lambda^{+})$. 
Note, that the function $\Gamma^{(L)} (a)$ should be linked to the \emph{phenomenological} 
entropy $S (a_1, \dots, a_d; V)$
of a nonequilibrium state with 
volume $V = L^n$ and the values of the other extensive coordinates $(a_1, a_2, \dots, a_d)$ 
via the Boltzmann's formula: 
\begin{equation*} 
S (a_1, \dots, a_d; V) = k_B \mathrm{ln} (c \, \Gamma^{(L)} (a)), 
\end{equation*}
where 
$k_B$ is the Boltzmann's constant, and 
$c$ is an arbitrary constant having the same units of measurement as the product 
$a_1 a_2 \dots a_d$ (the entropy in \emph{phenomenological} thermodynamics is defined up to an additive constant). 

At this point one can proceed as described above and consider the 
quasithermodynamic fluctuation theory over the points $(F (\cdot), \sigma (\cdot))$ 
in terms of the asymptotic expansions in the inverse rescaling parameter $\lambda^{-1} \to 0$. 
Without going into details, one can say the following.  
Denote the functions $(F (\cdot), \sigma (\cdot))$ 
corresponding to a point $\wt{\alpha} \in \wt{\Lambda}^{+}$ as 
$(F (\cdot; \wt{\alpha}), \sigma (\cdot; \wt{\alpha}))$. 
It follows from the construction of $\wt{\Lambda}^{+}$, that to every $\wt{\alpha} \in \wt{\Lambda}^{+}$ 
we can associate $L (\wt{\alpha}) := (\int d a \, F (a; \wt{\alpha}) )^{1/ n}$, and  
$\sigma (\cdot, \wt{\alpha}) \in \mathcal{Y}^{(L (\wt{\alpha}))}$. 
Assume that one can find a function $g_{\lambda} (\wt{\alpha})$ and 
a function  
$\sigma_{\lambda} (\cdot; \wt{\alpha}) \in \mathcal{Y}^{(\lambda^{1/ n} L (\wt{\alpha}))}$, 
such that the quantities $F (a, \wt{\alpha}, \lambda, \varepsilon)$ defined by 
\begin{equation*} 
F (a, \wt{\alpha}, \lambda, \varepsilon) := \Big( - k_B \frac{\delta}{\delta \sigma (a)} 
\mathrm{Tr} \big[ \wh{W}_{k_B, \varepsilon}^{(\lambda^{1/ n} L (\wt{\alpha}))} (\sigma (\cdot), g_{\lambda} (\wt{\alpha})) \big] 
\Big) \Big|_{\sigma (\cdot) = \sigma_{\lambda} (\cdot; \wt{\alpha})}
\end{equation*}
admit asymptotic expansions as $\lambda \to \infty$, 
\begin{equation*} 
\lambda^{-1} F (a, \wt{\alpha}, \lambda, \varepsilon) \simeq 
\sum_{l = 0}^{\infty} \lambda^{-l} F_{l} (a, \wt{\alpha}, \varepsilon), 
\end{equation*}
where the coefficients $F_{l} (a, \wt{\alpha}, \varepsilon)$, $l = 0, 1, 2, \dots$, 
are smooth enough and satisfy the condition that, for every 
fixed $\wt{\alpha}$ and $l$, 
$F_{l} (- , \wt{\alpha}, \varepsilon)$ has a weak limit as $\varepsilon \to 0$, and 
for the leading coefficient (corresponding to $l = 0$) this limit is $F (\cdot; \wt{\alpha})$. 
Consider the operators of \emph{flows} 
$\wh{\mathcal{I}}_{\varepsilon}^{(L)} (a; g)$ 
corresponding to $\wh{\mathcal{F}}_{\varepsilon}^{(L)} (a; g)$: 
\begin{equation*} 
\wh{\mathcal{I}}_{\varepsilon}^{(L)} (a; g) := 
\frac{\mathrm{i}}{\hbar} \Big[ 
\wh{\mathcal{H}}^{(L)} (g), \wh{\mathcal{F}}_{\varepsilon}^{(L)} (a; g)
\Big] = 
- 
\sum_{i = 1}^{d}
\wh{\mathcal{J}}_{i}^{(L)} (g)
\frac{\partial}{\partial a_i}
\wh{\mathcal{F}}_{\varepsilon}^{(L)} (a; g), 
\end{equation*}
where $\wh{\mathcal{H}}^{(L)} (g)$ is the second quantized Hamiltonian of the underlying multiparticle system, 
and $\wh{\mathcal{J}}_{i}^{(L)} (g)$ are the flow operators \eqref{eq:flow_operators} of the observables $\wh{\mathcal{E}}_{i}^{(L)} (g)$. 
The corresponding \emph{phenomenological} flows $\mathcal{I} (a; \wt{\alpha})$ 
can be perceived in analogy with \eqref{eq:flows_phenomen} as follows: 
\begin{equation*} 
\mathcal{I} (a; \wt{\alpha}) = 
\lim_{\varepsilon \to 0} 
\lim_{\lambda \to \infty} 
\bigg\lbrace 
\lambda^{-1} 
\mathrm{Tr} \Big[ \wh{W}_{k_B, \varepsilon}^{(L)} (\sigma (\cdot); g) 
\wh{\mathcal{I}}_{\varepsilon}^{(L)} (a; g)
\Big] \Big|_{L = \lambda^{1/ n} L (\wt{\alpha}), g = g_{\lambda} (\wt{\alpha}), \sigma (\cdot) = \sigma_{\lambda} (\cdot; \wt{\alpha})}
\bigg\rbrace. 
\end{equation*}
Expanding the definitions of the quantities in the square brackets, we obtain: 
\begin{multline*} 
\mathrm{Tr} \Big[ 
\wh{W}_{k_B, \varepsilon}^{(L)} (\sigma (\cdot); g) 
\wh{\mathcal{I}}_{\varepsilon}^{(L)} (a; g)
\Big] = 
- \sum_{i = 1}^{d} 
\frac{\partial}{\partial a_i} 
\mathrm{Tr} \Big\lbrace 
\exp \Big(- \frac{L^n}{k_B} \int d a' \, \sigma (a') \delta (a' - a) \Big) 
\times \\ \times 
\wh{\mathcal{J}}_{i}^{(L)} (g) L^n \prod_{j = 1}^{d} \delta_{\varepsilon} (a_j - \wh{\mathcal{E}}_{j}^{(L)} (g))
\Big\rbrace + o (\varepsilon). 
\end{multline*}
Performing the integration in the exponent, and taking into account, that 
for $\wt{\alpha} \in \wt{\Lambda}^{+}$ we have 
$\exp (- (L (\wt{\alpha}))^n \sigma (a; \wt{\alpha})/ k_B) = 
F (a; \wt{\alpha})/ ((L (\wt{\alpha}))^n \Gamma^{(L (\wt{\alpha}))} (a))$, we obtain: 
\begin{multline*} 
\mathrm{Tr} \Big[ 
\wh{W}_{k_B, \varepsilon}^{(L (\wt{\alpha}))} (\sigma (\cdot; \wt{\alpha}); g) 
\wh{\mathcal{I}}_{\varepsilon}^{(L (\wt{\alpha}))} (a; g)
\Big] = 
- \sum_{i = 1}^{d} \frac{\partial}{\partial a_i} \bigg( 
F (a; \wt{\alpha}) 
\times \\ \times 
\mathrm{Tr} 
\bigg\lbrace 
\wh{\mathcal{J}}_{i}^{(L (\wt{\alpha}))} (g) 
\frac{
\prod_{j = 1}^{d} \delta_{\varepsilon} (a_j - \wh{\mathcal{E}}_{j}^{(L (\wt{\alpha}))} (g))
}{\Gamma^{(L (\wt{\alpha}))} (a)} 
\bigg\rbrace
\bigg) + o (\varepsilon). 
\end{multline*}
Therefore, the expression for the phenomenological flows reduces to 
\begin{equation*} 
\mathcal{I} (a; \wt{\alpha}) = 
- \sum_{i = 1}^{d} \frac{\partial}{\partial a_i} 
\Big( 
F (a; \wt{\alpha}) u_{i} (a; \wt{\alpha})
\Big), 
\end{equation*}
where 
\begin{equation*} 
u_{i} (a; \wt{\alpha}) := 
\lim_{\varepsilon \to 0}
\lim_{\lambda \to \infty} 
\bigg(
\lambda^{-1} 
\mathrm{Tr} 
\bigg\lbrace 
\wh{\mathcal{J}}_{i}^{(L)} (g) 
\frac{
\prod_{j = 1}^{d} \delta_{\varepsilon} (a_j - \wh{\mathcal{E}}_{j}^{(L)} (g))
}{\Gamma^{(L)} (a)} 
\bigg\rbrace \bigg|_{L = L_{\lambda} (\wt{\alpha}), g = g_{\lambda} (\wt{\alpha})}
\bigg), 
\end{equation*}
for $i = 1, 2, \dots, d$. 
Like in the beginning of this section, 
the evolution of the system is described as a curve $\wt{\gamma} = \lbrace \wt{\alpha}^t \rbrace_t$ 
on $\wt{\Lambda}^{+}$. 
There exists an equation describing this curve, the derivation of 
which is totally similar to the derivation of \eqref{eq:Onsager}, see \cite{ZubarevMorozovRopke_vol2} for details. 
Under the assumptions of the ``smallness of time-memory effects'' and 
``smallness of spatial gradients'', 
one arrives at the \emph{generalized Fokker-Planck equation}: 
\begin{equation*} 
\frac{\partial}{\partial t} F (a; \wt{\alpha}^t) 
+ \sum_{i = 1}^{d} 
\frac{\partial}{\partial a_i} \Big( F (a; \wt{\alpha}^t) u_i (a; \wt{\alpha}^t)\Big) 
- \sum_{i, j = 1}^{d} 
\bigg( 
\frac{\partial}{\partial a_i} 
\mathcal{K}_{i, j} (a; \wt{\alpha}^{t}) \frac{\partial}{\partial a_j} \bigg)
\frac{F (a; \wt{\alpha}^{t})}{\Gamma^{L (\wt{\alpha}^{t})} (a)}
= 0, 
\end{equation*}
where 
$\mathcal{K}_{i, j} (a; \wt{\alpha}^t)$ are some coefficients 
(their role is similar to the Onsager coefficients $L_{i, j} (\alpha^t)$ in \eqref{eq:Onsager}), 
and $\Gamma^{(L)} (a)$ is the function defining the limiting manifold $\wt{\Lambda}^{+}$ 
(i.e. the \emph{phenomenological} statistical weight, see \eqref{eq:Lambda_square}). 
It can be more convenient to write this equation in the following form: 
\begin{equation}
\label{eq:Fokker_Planck_diffusion}
\frac{\partial}{\partial t} F (a; \wt{\alpha}^t) 
+ \sum_{i = 1}^{d} 
\frac{\partial}{\partial a_i} \Big( F (a; \wt{\alpha}^t) v_i (a; \wt{\alpha}^t)\Big) 
- \sum_{i, j = 1}^{d} 
\bigg( 
\frac{\partial}{\partial a_i} 
\mathcal{D}_{i, j} (a; \wt{\alpha}^{t}) \frac{\partial}{\partial a_j} \bigg)
F (a; \wt{\alpha}^{t})
= 0, 
\end{equation}
where 
\begin{equation*} 
\mathcal{D}_{i, j} (a; \wt{\alpha}) := 
\frac{\mathcal{K}_{i, j} (a; \wt{\alpha})}{\Gamma^{(L (\wt{\alpha}))} (a)} 
\end{equation*}
are the \emph{diffusion coefficients} ($i, j = 1, 2, \dots, d$, $\wt{\alpha} \in \wt{\Lambda}^{+}$, $a \in \mathbb{R}^{d}$), and 
\begin{equation*}
v_{i} (a; \wt{\alpha}) := u_{i} (a; \wt{\alpha}) + 
\sum_{j = 1}^{d}
\frac{\mathcal{K}_{i, j} (a; \wt{\alpha})}{\Gamma^{(L (\wt{\alpha}))} (a)} 
\Big( \frac{\partial}{\partial a_j} \mathrm{ln} \Gamma^{(L (\wt{\alpha}))} (a) \Big), 
\end{equation*}
are the \emph{coefficients of drift}
($i = 1, 2, \dots, d$, $\wt{\alpha} \in \wt{\Lambda}^{+}$, $a \in \mathbb{R}^{d}$). 
If we denote, for every $i = 1, 2, \dots, d$, 
\begin{equation*}
\beta_{i}^{(L)} (a) := k_B \frac{\partial}{\partial a_i} \mathrm{ln} \Gamma^{(L)} (a), 
\end{equation*}
(these are the analogues of the inverse absolute temperature and other intensive thermodynamic quantities 
in the equilibrium thermodynamics), then we obtain 
\begin{equation*} 
v_{i} (a; \wt{\alpha}) = u_{i} (a; \wt{\alpha}) + 
\sum_{j = 1}^{d} \mathcal{D}_{i, j} (a; \wt{\alpha}) \beta_{j}^{(L (\wt{\alpha}))} (a), 
\end{equation*}
for $i = 1, 2, \dots, d$, $\wt{\alpha} \in \wt{\Lambda}^{+}$, $a = (a_1, a_2, \dots, a_d) \in \mathbb{R}^{d}$. 
The equation \eqref{eq:Fokker_Planck_diffusion} is more familiar in the phenomenological physics, 
since in practice, $L (\wt{\alpha}^{t})$ is quite often fixed along the evolution, and 
the coefficients depend very little on $a$ and $\wt{\alpha}$ and can be determined empirically 
(recall, for example, the \emph{Fourier's law} for the heat transfer, or the \emph{Fick's law} for diffusion). 

The equation \eqref{eq:Fokker_Planck_diffusion} can be perceived as a ``more precise'' description 
of a physical system, than the one given by \eqref{eq:Onsager}. 
Suppose $L (\wt{\alpha}^{t}) = \mathrm{const}$ along the evolution. 
Then it is more convenient to introduce 
$f (a; \wt{\alpha}) := (L (\wt{\alpha}))^{-n} F (a; \wt{\alpha})$, which is normalized as $\int da \, f (a; \wt{\alpha}) = 1$. 
Instead of a curve $\lbrace \alpha^t \rbrace_{t} \subset \Lambda^{+}$, 
we now have a curve $\lbrace \wt{\alpha}^{t} \rbrace_t \subset \wt{\Lambda}^{+}$. 
This function describes the \emph{density} of distribution of ``fluctuations'' of $E_1, E_2, \dots, E_d$ 
(recall, that $\Lambda^{+}$ is a Lagrangian manifold 
embedded in a phase space with extensive coordinates $(x_1, \dots, x_{d + 1}) = (E_1, \dots, E_d, V)$ and 
intensive coordinates $(y_1, \dots, y_{d + 1}) = (\beta_1, \dots, \beta_d, \wt{p})$, 
the symplectic structure is $\omega^{+} = \sum_{i = 1}^{d + 1} d y_i \wedge d x_i$).  
We have a pair of \emph{consistency} conditions between the two descriptions. 
The first one is that the \emph{entropy} $S (E_1, \dots, E_d, L^n) = 
\int \sum_{i = 1}^{d} \beta_i d E_i$, 
$L|_{\gamma} = V^{1/ n} = \mathrm{const}$, 
corresponding to the Lagrangian manifold $\Lambda^{+}$, 
is linked to the phenomenological statistical weight $\Gamma^{(L)} (a)$, $a = (a_1, \dots, a_d)$ 
associated with $\wt{\Lambda}^{+}$, via the \emph{Boltzmann's formula}: 
\begin{equation*} 
S (E_1, \dots, E_d, L^{n}) = k_B \mathrm{ln} ( c \Gamma^{(L)} (E_1, \dots, E_d)), 
\end{equation*}
where $k_B$ is the Boltzmann's constant, 
and $c > 0$ is a constant with the same units of measurement as $(E_{1} E_{2} \dots E_{d})^{-1}$. 
The second one is a condition on the mathematical expectation: 
\begin{equation*} 
E_i (\alpha^{t}) = \int d a\, a_i f (a; \wt{\alpha}^{t}), 
\end{equation*}
where $i = 1, 2, \dots, d$, 
(note, that this equality becomes approximate, once we use the Onsager's equations \eqref{eq:Onsager} 
and the generalized Fokker-Planck equations \eqref{eq:Fokker_Planck_diffusion}). 
If we wish to consider \eqref{eq:Fokker_Planck_diffusion} with an initial condition 
$F (a, \wt{\alpha}^t)|_{t = 0} = L^{n} f_0 (a)$, 
then for the rest we are \emph{not} restricted in a choice of $f_0 (\cdot)$ 
(i.e. the starting point on $\wt{\Lambda}^{+}$ is for the rest arbitrary). 
In particular, consider the cumulants (assuming they exist): 
\begin{equation*} 
\wt{C}_{N} (t) := 
\bigg[
\Big( - \mathrm{i} k_{B} \frac{\partial}{\partial b} \Big)^N
\int d a\, f(a, \wt{\alpha}^t) \exp \Big( \frac{\mathrm{i}}{k_B} \sum_{i = 1}^{d} b_i a_i \Big)
\bigg] \bigg|_{b = \bar 0}, 
\end{equation*}
where $b = (b_1, \dots, b_d)$, 
$N = (n_1, \dots, n_d) \in \mathbb{Z}_{\geqslant 0}^{d}$, 
$\bar 0 = (0, \dots, 0) \in \mathbb{R}^d$, $N \not = \bar 0$,  
and we use a standard notation for multi-indices: 
$(- \mathrm{i} k_B \partial/ \partial b)^{N} := 
\prod_{j = 1}^{d} (- \mathrm{i} k_B \partial/ \partial b_j)^{n_j}$.
The second consistency condition mentioned is just 
\begin{equation*} 
\wt{C}_{N} (t) = \lim_{\lambda \to \infty} \lambda^{-1} C_{(N, 0)}^{(\lambda)} (\alpha^t), 
\end{equation*}
if $|N| := \sum_{i = 1}^{d} n_i = 1$, where $N = (n_1, \dots, n_d)$, and 
$(N, 0) := (n_1, \dots, n_d, 0)$. On the other hand, this equality 
is not required in case where $|N| \geqslant 2$, 
i.e. the cumulants $\wt{C}_{N} (t)$ should be perceived as 
derivatives of the action function corresponding to \emph{different} 
Lagrangian manifolds $\mathcal{M}_t$ ``attached'' to the points $\alpha^t \in \Lambda^{+}$. 
More precisely, we need only the \emph{germs} of $\mathcal{M}_t$ of these Lagrangian manifolds 
(which are termed the \emph{microlagrangian manifolds}). 
Note, that 
this construction is similar to the construction used in the method of canonical 
operator with a complex phase \cite{DubnovMaslovNazaikinskii, Maslov_op_meth}. 

How do $\mathcal{M}_{t}$ change with time? 
More generally, we would like to consider a ``deformation'' of the linear fluctuation theory 
with respect to the rescaling parameter $\lambda \to \infty$, 
i.e. to introduce the functions $\wt{C}_{N}^{(\lambda)} (t)$ generalizing $\wt{C}_{N} (t)$, 
$N \in \mathbb{Z}_{\geqslant 0}^{d}$, $N \not = \bar 0$. 
How does the collection $\lbrace \wt{C}_{N}^{(\lambda)} (t) \rbrace_{N}$ change with time? 
The main idea which allows to derive these equations is as follows: 
let us consider the limit $\lambda^{-1} \to 0$ in analogy with the \emph{semiclassical limit} $h \to 0$ of quantum mechanics 
(where $h$ is the small parameter of the transition corresponding to the Planck constant $\hbar$). 

Suppose we have a \emph{classical} 
mechanical system with $n$ degrees of freedom $q = (q_1, q_2, \dots, q_n) \in\mathbb{R}^{n}$ 
described by a Hamiltonian $H (p, q)$, where $p = (p_1, p_2, \dots, p_n)$ are the 
canonically conjugate momenta corresponding to $q$. 
Assume that $H$ is a smooth function, say, from the Schwartz space $\mathcal{S} (\mathbb{R}_{q, p}^{2 n})$ 
of rapidly decaying functions at infinity. 
A \emph{semiclassical} analogue of this system is described 
by a Hamiltonian $H_{h} (q, p)$ which depends on the small parameter $h \to 0$ of semiclassical approximation, 
$H_{h} (q, p) = H (q, p) + \sum_{s = 1}^{r_0} ((- \mathrm{i} h)^{s}/ s!) H^{(s)} (q, p)$, 
where $r_0$ is a positive integer, and $H^{(s)} \in \mathcal{S} (\mathbb{R}_{q, p}^{2 n})$, 
$s = 1, 2, \dots, r_{0}$.

A semiclassical wave function $\psi_{h}^{t} (q)$ satisfies modulo $O (h^{\infty})$ the Schr\"odinger 
equation with a small parameter $h$ in front of partial derivatives: 
\begin{equation*} 
\mathrm{i} h \frac{\partial}{\partial t} \psi_{h}^{t} (q) = 
H_{h} \Big( 
- \mathrm{i} h \frac{\partial}{\partial q}, q
\Big) \psi_{h}^{t} (q) + O (h^{\infty}), 
\end{equation*}
where 
one assumes the Weyl quantization on the right-hand side, and 
one defines $O (h^{\infty})$ as follows: 
$\psi_{h}^{t} (q) = O (h^{\infty})$ if and only if for all $k = 1, 2, 3, \dots$, 
there exists $C_k^t > 0$, such that $\sup_{q \in \mathbb{R}^n} |\psi_{h}^{t} (q)| \leqslant C_k^t h^k$. 
For the semiclassical wave functions there exists a limit 
\begin{equation*} 
\rho^t (q) := \lim_{h \to 0} |\psi_{h}^{t} (q)|^{2}, 
\end{equation*}
(invoke the Bohr's correspondence principle). 
The main idea is to try to perceive $f (a; \wt{\alpha}^{t}) = L^{-n} F (a, \wt{\alpha}^t)$ corresponding to 
the equation \eqref{eq:Fokker_Planck_diffusion} in analogy with $\rho^t (q)$. 
The small parameter $h \to 0$ should correspond to $\lambda^{-1} \to 0$. 
Consider first the Wigner's \emph{quasiprobability} function associated with $\psi_{h}^{t} (q)$: 
\begin{equation*} 
\rho_{h}^{t} (q, p) := \frac{1}{(2 \pi h)^{n}} 
\int d q' \, 
\exp \Big( - \frac{\mathrm{i}}{h} p q \Big) 
\bar \psi_{h}^{t} \Big( q - \frac{q'}{2} \Big) 
\psi_{h}^{t} \Big( q + \frac{q'}{2} \Big),
\end{equation*} 
where the bar denotes the complex conjugation, and $(q, p) \in \mathbb{R}_{q, p}^{2 n}$. 
It follows, that 
\begin{equation*} 
\rho^t (q) = \int d p\, 
\lim_{h \to 0} \rho_{h}^{t} (q, p), 
\end{equation*}
where the limit under the integral on the right-hand side is taken in the weak sense. 
Note, that $\rho_h^t (q, p)$ is real, 
$\rho_h^t (q, p) = \bar \rho_{h}^{t} (q, p)$, but, in general, 
does \emph{not} need to be positively defined, 
and satisfies only the normalization condition 
$\int d p d q\, \rho_{h}^t (q, p) = 1$.  
The classical limit $\rho^t (q, p)$, on the other hand, satisfies $\rho^t (q, p) \geqslant 0$. 
Consider now $N$ semiclassical particles with an interaction 
between a pair of particles concentrated in $(q, p)$ and $(q', p')$ described by 
\begin{equation*} 
V_h (q, p; q', p') := V (q, p; q', p') + \sum_{s = 1}^{r_1} \frac{(- \mathrm{i} h)^{s}}{s!} 
V^{(s)} (q, p; q', p'), 
\end{equation*}
where $r_1$ is a positive integer, and $V$ and $V_s$, $s = 1, 2, 3, \dots$, 
are elements of $\mathcal{S} (\mathbb{R}_{q, p}^{2 n} \times \mathbb{R}_{q, p}^{2 n})$ 
(i.e. Swartz functions on a 2-particle phase space).  
To describe the \emph{kinetics} of this system, one introduces 
the 1-particle density function $R_{N, h}^t (q, p)$ (the analogue of $\rho_{h}^t (q, p)$) 
normalized on the number of particles, 
\begin{equation*} 
\int_{\mathbb{R}^{2 n}} d q d p \, R_{N, h}^t (q, p) = N, 
\end{equation*}
and considers a limit $N \to \infty$, $h \to 0$, in such a way that 
\begin{equation*}
N h^{n} = \varkappa = \mathrm{const} > 0. 
\end{equation*}
Assume that we adjust the ``geometry'' of the configuration space of the system 
(this can be the radius $L$ of a $n$-dimensional torus, if the extracted system is confined on a torus), 
as well as the other parameters $g = (g_1, g_2, \dots, g_m)$ describing the ``nature'' of the 
system 
(for instance, the radius of interaction or a parameter describing the external field)
in such a way, that for the weak limit $\rho^t (q, p)$ of 
$R_{N, h}^{t} (q, p)/ N$ (where $N \to \infty$ and $N h^n = \varkappa > 0$ is fixed), 
one obtains an equation of the shape 
\begin{equation} 
\label{eq:kinetic}
\frac{\partial}{\partial t} \rho^t (q, p) + 
\big\lbrace H [\rho^t] (q, p; \varkappa), \rho^t (q, p) \big\rbrace - J [\rho^t] (q, p; \varkappa) = 0, 
\end{equation}
where $\lbrace -, - \rbrace$ denotes the canonical Poisson bracket on the 1-particle phase space, 
with coordinates 
$q = (q_1, q_2, \dots, q_n)$ and momenta $p = (p_1, p_2, \dots, p_n)$, 
$\lbrace p_i, q_j \rbrace = \delta_{i, j}$, $i, j \in [n]$, 
the function $H [\rho^t] (q, p; \varkappa)$ is the \emph{selfconsistent Hamiltonian}, 
and $J [\rho^t] (q, p; \varkappa)$ is the \emph{collision integral} \cite{ZubarevMorozovRopke_vol1}. 
A usual way to analyse this equation is via an asymptotic expansion 
of the collision integral in $\varkappa \to 0$ and by constructing the BBGKY chain of equations. 
Observe, that if we integrate \eqref{eq:kinetic} over $p$, then this yields an expression in which 
the term corresponding to the Poisson bracket can be associated with 
the first sum in \eqref{eq:Fokker_Planck_diffusion} (the sum over one index), and 
the term corresponding to the collision integral can be associated with 
the second sum in \eqref{eq:Fokker_Planck_diffusion} (the sum over two indices).   

To make the analogy more explicit, let us look at our thermodynamic system and consider 
along with $\wh{\mathcal{E}}_{1}^{(L)} (g), 
\dots, \wh{\mathcal{E}}_{d}^{(L)} (g)$, 
the flows  
$\wh{\mathcal{J}}_{1}^{(L)} (g), 
\dots, \wh{\mathcal{J}}_{d}^{(L)} (g)$ 
defined in \eqref{eq:flow_operators} 
(recall, that $L$ denotes the radius of the $n$-dimensional torus on which we consider the system, 
and $g = (g_1, g_2, \dots, g_m)$ is a collection of parameters describing the interaction and the external field). 
Recall, that it is also assumed (for simplicity), that 
$\lbrace \wh{\mathcal{E}}_{i}^{(L)} (g) \rbrace_{i = 1}^{d}$ mutually commute, and that  
$\lbrace \wh{\mathcal{J}}_{i}^{(L)} (g)\rbrace_{i = 1}^{d}$ mutually commute. 
Look first at the analogue of \eqref{eq:Onsager} (we assume that the volume $V = L^n$ is fixed along the 
evolution). 
One needs to consider the phase space $\mathbb{R}^{4 d + 2}$ with 
``coordinates'' $(x_1, \dots, x_d, x_{d+1}, \dots, x_{2 d}, V)$ and 
``momenta'' $(y_1, \dots, y_d, y_{d + 1}, \dots, y_{2 d}, \wt{p})$, 
where for $s = 1, 2, \dots, d$, 
$x_{s}$ corresponds to $\wh{\mathcal{E}}_{s}^{(L)} (g)$, 
$x_{d + s}$ corresponds to $\wh{\mathcal{J}}_{d + s}^{(L)} (g)$, 
$y_{s}$ corresponds to the ``inverse temperature'' conjugate to $\wh{\mathcal{E}}_{s}^{(L)} (g)$, and 
$y_{d + s}$ corresponds to the ``inverse temperature'' conjugate to $\wh{\mathcal{J}}_{s}^{(L)} (g)$. 
It is convenient to denote $x_{2 d + 1} = V$, and $y_{2 d + 1} = \wt{p}$ 
(the intensive quantity ``pressure over absolute temperature'' 
associated with the volume $V$). 
The symplectic structure is $\omega^{\#} := \sum_{i = 1}^{2 d + 1} d y_i \wedge d x_i$, and 
the evolution is described as a curve $\gamma = \lbrace \alpha^t \rbrace_{t} \subset \Lambda^{\#}$ 
on a Lagrangian manifold $\Lambda^{\#} \subset \mathbb{R}^{4 d + 2}$, where $t$ is time. 
Denote $x_i (\alpha)$ and $y_i (\alpha)$, $i = 1, 2, \dots, 2 d + 1$, the coordinates of a point $\alpha \in \Lambda^{\#}$ 
acquired in the ambient phase space. 
The curve satisfies the following system of equations (in analogy with \eqref{eq:Onsager}): 
\begin{equation} 
\label{eq:Onsager_double}
\frac{\partial x_{i} (\alpha^t)}{\partial t} = 
J_{i} (\alpha^t) + \sum_{j = 1}^{2 d} L_{i, j} (\alpha^t) y_j (\alpha^t), 
\end{equation}
where $i = 1, 2, \dots, 2 d$, and $J_i (\alpha)$ and $L_{i, j} (\alpha)$, $i, j \in [2 d]$, 
are given functions on $\Lambda^{\#}$ 
(the flows and the Onsager coefficients). 
These functions can be perceived as a limit $\lambda \to \infty$ of more complicated quantities associated with rescaled systems. 
In particular, for $J_{s} (\alpha)$, $s \in [d]$, one needs to consider
the flows of $\wh{\mathcal{E}}_{i}^{(L)} (g)$ given by the operators \eqref{eq:flow_operators} and for 
$J_{d + s} (\alpha)$ one needs to consider the flows of the flows $\wh{\mathcal{J}}_{s}^{(L)} (g)$, 
\begin{equation*} 
\wh{\mathcal{J}}_{d + s}^{(L)} (g) := \frac{\mathrm{i}}{\hbar} \Big[ 
\wh{\mathcal{H}}^{(L)} (g), \wh{\mathcal{J}}_{s}^{(L)} (g)
\Big] = 
\frac{\mathrm{i}}{\hbar} \Big[ 
\wh{\mathcal{H}}^{(L)} (g), 
\frac{\mathrm{i}}{\hbar} \Big[ 
\wh{\mathcal{H}}^{(L)} (g), 
\wh{\mathcal{E}}_{s}^{(L)} (g)
\Big] \Big], 
\end{equation*}
where $s \in [d]$, 
(we have a \emph{family} of underlying systems parametrized by $L$ and $g$). 
If we denote $\wh{\mathcal{E}}_{d + s}^{(L)} (g) := \wh{\mathcal{J}}_{s}^{(L)} (g)$, $s \in [d]$, 
then the space of relevant ensembles is described by the Gibbs's operators 
\begin{equation*} 
\wh{w}_{k_B}^{(L)} (\beta; g) := \frac{1}{\mathcal{Z}_{k_B}^{(L)} (\beta; g)} 
\exp \Big( 
- \frac{1}{k_B} \sum_{i = 1}^{2 d} \beta_i \wh{\mathcal{E}}_{i}^{(L)} (g)
\Big), 
\end{equation*}
where $\beta = (\beta_1, \dots, \beta_{2 d})$, and 
$\mathcal{Z}_{k_B}^{(L)} (\beta; g)$ is determined by the normalization 
condition $\mathrm{Tr} \wh{w}_{k_B}^{(L)} (\beta; g) = 1$ (assuming the trace exists). 
For every $\alpha \in \Lambda^{\#}$ there exist functions 
$\beta_{\lambda} (\alpha)$ and $g_{\lambda} (\alpha)$ 
depending on the large rescaling parameter $\lambda \to \infty$ in such a way, that 
\begin{equation*} 
J_{i} (\alpha) = \lim_{\lambda \to \infty} 
\bigg\lbrace 
\lambda^{-1} 
\mathrm{Tr} 
\Big[ \wh{\mathcal{J}}_{i}^{(L)} (g) \wh{w}_{k_B}^{(L)} (\beta; g) 
\Big] \Big|_{L = (\lambda x_{2 d + 1} (\alpha))^{1/ n}, \beta = \beta_{\lambda} (\alpha), g = g_{\lambda} (\alpha)} 
\bigg\rbrace, 
\end{equation*}
where $i \in [2 d]$. The coefficients $L_{i, j} (\alpha)$ can be perceived in a similar way 
as a limit of more complicated expressions (involving, for example, the projection operators of Mori \cite{Mori} and 
Zwanzig \cite{Zwanzig}). 
It is important to point out, that the system \eqref{eq:Onsager_double} 
is valid in the approximation of ``small spatial gradients''. 
To compute the ``flows of the flows'', one applies the commutator 
$(\mathrm{i}/\hbar) [\wh{\mathcal{H}}^{(L)} (g), -]$ twice, and, therefore, we should put 
\begin{equation} 
\label{eq:flow_flow_zero}
J_{d + s} (\alpha^{t}) = 0, 
\end{equation}
where $s \in [d]$, along the evolution curve 
$\gamma = \lbrace \alpha^t \rbrace_t \subset \Lambda^{\#}$, see \cite{Zwanzig} for details. 
The condition \eqref{eq:flow_flow_zero} simply restricts the possible curve $\gamma$ and implies, 
that the Onsager's matrix $\| L_{i, j} (\alpha^t) \|_{i, j = 1}^{2 d}$ cannot be arbitrary.

Now, if we wish to describe the ``fluctuations'' of the values of $x_i (\alpha^t)$, $y_i (\alpha^t)$, $i \in [2 d]$, 
corresponding to a point $\alpha^t \in \gamma \subset \Lambda^{\#}$ of the evolution curve, 
then we should introduce a ``phase space'' formed by the pairs 
$(\wt{F} (x), \wt{\sigma} (x))$ ($\wt{F}$ is a distribution, and $\wt{\sigma}$ is a function), 
where $x = (x_1, x_2, \dots, x_{2 d}) \in \mathbb{R}^{2 d}$. 
For the points $\alpha \in \Lambda^{\#}$, we have 
\begin{equation*} 
y_i (\alpha) = \frac{\partial}{\partial x_i} \wt{S} (x_1, \dots, x_{2 d}; V (\alpha)) \bigg|_{x_{j} = x_{j} (\alpha), j \in [2 d]}, 
\end{equation*}
where $i \in [2 d]$, and $\wt{S} (x_1, \dots, x_{2 d}; V)$ is 
the \emph{phenomenological} nonequilibrium entropy (see \cite{Prigogine}) 
(note that this is a function of extensive thermodynamic coordinates 
$(x_1, \dots, x_d)$, their flows $(x_{d + 1}, \dots, x_{2 d})$, and the volume $V$). 
Define the \emph{phenomenological} statistical weight $\wt{\Gamma}^{(L)} (x_1, \dots, x_{2 d})$ 
from the Boltzmann's formula: 
\begin{equation*}
\wt{S} (x_1, \dots, x_{2 d}; L^n) = k_B \mathrm{ln} (c \wt{\Gamma}^{(L)} (x_1, \dots, x_{2 d})), 
\end{equation*}
where $c > 0$ is a constant with the same units of measurement as $(x_1 \dots x_{2 d})^{-1}$, 
and construct a ``huge'' Lagrangian manifold $\wt{\Lambda}^{\#}$ like \eqref{eq:Lambda_square}, 
\begin{equation} 
\label{eq:Lambda_huge}
\wt{\Lambda}^{\#} := 
\bigcup_{L > 0}
\Big\lbrace (\wt{F} (\cdot), \wt{\sigma} (\cdot)) \,|\, 
\wt{\sigma} (\cdot) \in \wt{\mathcal{Y}}^{(L)} \,\&\, 
\forall x: 
\wt{F} (x) =  L^n \exp \Big( - \frac{L^n}{k_B} \wt{\sigma} (x) \Big) \wt{\Gamma}^{(L)} (x)
\Big\rbrace, 
\end{equation}
where $x = (x_1, x_2, \dots, x_{2 d})$, and 
$\wt{\mathcal{Y}}^{(L)} := \lbrace \wt{\sigma} \,|\, \int d x\, \wt{\Gamma}^{(L)} (x) \exp(- L^n \wt{\sigma} (x)/ k_B) = 1 \rbrace$. 
The difference with the formula \eqref{eq:Lambda_square} is only that we have $2 d$ arguments in the functions, in place of $d$. 
The evolution of the system is a curve $\wt{\gamma} = \lbrace \wt{\alpha}^t \rbrace_t \subset \wt{\Lambda}^{\#}$ 
described by a system of equations similar to \eqref{eq:Fokker_Planck_diffusion} and \eqref{eq:Onsager_double} 
with a condition similar to \eqref{eq:flow_flow_zero}. 

Denote $( \wt{F} (x; \wt{\alpha}), \wt{\sigma} (x; \wt{\alpha}) )$ the coordinates of a point $\wt{\alpha} \in \wt{\Lambda}^{\#}$ 
stemming from the phase space. 
Observe, that $\wh{\mathcal{J}}_{s}^{(L)} (g) = \wh{\mathcal{E}}_{d + s}^{(L)} (g)$, $s \in [d]$, 
is present in the product of delta functions 
$\prod_{i = 1}^{2 d} \delta (x_i - \wh{\mathcal{E}}_{i}^{(L)} (g))$ corresponding to the analogue of the microcanonical distribution in this case. 
Therefore, computing $u_{i} (x; \wt{\alpha})$ 
for the generalized Fokker-Planck equation 
(where $i \in [2d]$, $x = (x_1, \dots, x_{2 d})$, $\wt{\alpha} \in \wt{\Lambda}^{\#}$), 
we obtain $u_{s} (x; \wt{\alpha}) = x_{d + s}$, $s \in [d]$. 
The quantities $u_{d + s} (x; \wt{\alpha})$ correspond to applying the commutator 
$(\mathrm{i}/ \hbar) [ \wh{\mathcal{H}}^{(L)} (g), -]$ twice (i.e. ``flows of the flows''), 
and in the approximation of ``small gradients'' we should put them to zero \cite{Zwanzig}. 
In the end, this yields the following equation for $\wt{F} (x; \wt{\alpha}^t)$: 
\begin{multline*} 
\frac{\partial \wt{F} (x; \wt{\alpha}^{t})}{\partial t} 
+ \sum_{s = 1}^{d} x_{d + s} \frac{\partial \wt{F} (x; \wt{\alpha}^t)}{\partial x_s} 
+ \sum_{i = 1}^{2 d} \frac{\partial}{\partial x_i} 
\Big( 
\wt{F} (x; \wt{\alpha}^t) 
\sum_{j = 1}^{2 d} \mathcal{\wt{D}}_{i, j} (x; \wt{\alpha}^t) \, 
\beta_j^{(L (\wt{\alpha}^t))} (x)
\Big)  
- \\ -
\sum_{i, j = 1}^{2 d} 
\bigg( 
\frac{\partial}{\partial x_i} \wt{\mathcal{D}}_{i, j} (x; \wt{\alpha}^t) \frac{\partial}{\partial x_j} 
\bigg)
\wt{F} (x; \wt{\alpha}^t) = 0, 
\end{multline*}
where $\wt{\mathcal{D}}_{i, j} (x; \wt{\alpha})$ are the analogue of the diffusion coefficients 
in \eqref{eq:Fokker_Planck_diffusion}, 
and $\wt{\beta}_j^{(L)} (x)$ are defined as 
$\wt{\beta}_j^{(L)} (x) = (k_B \partial/ \partial x_j ) \mathrm{ln} \wt{\Gamma}^{(L)} (x)$, $j \in [2 d]$. 
The vector field on $\mathbb{R}_{x}^{2 d}$ with the components 
$Y_i (x; \wt{\alpha}) := \sum_{j = 1}^{2 d} \mathcal{\wt{D}}_{i, j} (x; \wt{\alpha}^t) \, 
\beta_j^{(L (\wt{\alpha}^t))} (x)$, $i \in [2d]$, describes a contribution to the total flow 
associated with $\wh{\mathcal{E}}_{i}^{(L)} (g)$ induced by a deviation of the system from a thermodynamic equilibrium. 
Let us assume that the divergence of this vector field vanishes: 
\begin{equation} 
\label{eq:divergence_zero}
\sum_{i = 1}^{2 d} \frac{\partial}{\partial x_i} Y_i (x; \wt{\alpha}) = 0. 
\end{equation}
Then, for every $\wt{\alpha} \in \wt{\Lambda}^{\#}$, we can find a function $\Phi (x; \wt{\alpha})$ such that 
$Y_{i} (x; \wt{\alpha}) = \partial \Phi(x; \wt{\alpha})/ \partial x_i$, $i \in [2 d]$, $x \in \mathbb{R}^{2 d}$. 
Furthermore, let us represent this derivative as 
$\partial \Phi(x; \wt{\alpha})/ \partial x_i = \sum_{j = 1}^{2 d} J_{i, j} \partial V (x; \wt{\alpha})/ \partial x_j$, 
where $J_{i, j} = \delta_{i, j - d} - \delta_{i - d, j}$, $\delta$ is the Kronecker symbol.  
The matrix $J = \| J_{i, j} \|_{i, j = 1}^{2 d}$ satisfies $J^{-1} = J^{T} = - J$, where $(\cdot)^{T}$ denotes the transposed matrix. 
Then we have $\partial V (x; \wt{\alpha})/ \partial x_i = \sum_{j = 1}^{2 d} (J^{-1})_{i, j} \partial \Phi (x)/ \partial x_j$, 
and the function $V (-; \wt{\alpha})$ exists since 
$\sum_{i = 1}^{2 d} \partial/ \partial x_i \big( \sum_{j = 1}^{2 d} (J^{-1})_{i, j} \partial \Phi (x)/ \partial x_j \big) = 0$ 
due to $J^{T} = - J$. 
So one obtains: 
\begin{equation*} 
Y_s (x; \wt{\alpha}) = \frac{\partial V (x; \wt{\alpha})}{\partial x_{d + s}}, \quad 
Y_{d + s} (x; \wt{\alpha}) = - \frac{\partial V (x; \wt{\alpha})}{\partial x_{s}},  
\end{equation*}
for $s \in [d]$. 
Substituting this into the generalized Fokker-Planck equation for $\wt{F} (x; \wt{\alpha}^{t})$, we obtain: 
\begin{equation} 
\label{eq:Fokker_Planck_phase_space}
\frac{\partial \wt{F} (x; \wt{\alpha}^t)}{\partial t} + 
\sum_{s = 1}^{d} \bigg( 
\frac{\partial H (x; \wt{\alpha}^{t})}{\partial x_{d + s}} 
\frac{\partial \wt{F} (x; \wt{\alpha}^{t})}{\partial x_{s}} - 
\frac{\partial H (x; \wt{\alpha}^{t})}{\partial x_{s}} 
\frac{\partial \wt{F} (x; \wt{\alpha}^{t})}{\partial x_{d + s}} 
\bigg) = I (x; \wt{\alpha}^t), 
\end{equation}
where $I (x, \wt{\alpha}) := \sum_{i, j = 1}^{2 d} ( (\partial/ \partial x_i) \mathcal{D}_{i, j} (x; \wt{\alpha}) 
(\partial/ \partial x_{j}) ) \wt{F} (x; \wt{\alpha})$, and 
\begin{equation*} 
H (x; \wt{\alpha}) := \frac{1}{2} \sum_{s = 1}^{d} x_{d + s}^{2} + V (x; \wt{\alpha}),  
\end{equation*}
where $x = (x_1, \dots, x_{2 d}) \in \mathbb{R}^{2 d}$, and $\wt{\alpha} \in \wt{\Lambda}^{\#}$. 
It is suggested to perceive $H (x; \wt{\alpha})$ as an analogue of a \emph{self-consistent Hamiltonian} in mechanics 
(i.e. $V (x; \wt{\alpha})$ corresponds to the \emph{dressed potential}), 
and the quantity $I (x; \wt{\alpha})$ as an analogue of the \emph{collision integral}. 
The assumption \eqref{eq:divergence_zero} about the divergence of the 
vector field describing the thermodynamic flows 
is basically an assumption about applicability of a self-consistent picture of description, 
which, of course, depends on a ``successful'' choice of the collection of the basis quantities 
$\lbrace \wh{\mathcal{E}}_{i}^{(L)} (g) \rbrace_{i = 1}^{d}$. 

The basic idea is now as follows: 
let us perceive $\wt{F} (x; \wt{\alpha})$ in analogy with the weak limit of the Wigner's 
quasiprobability function $\rho_{h} (q, p)$ in quantum mechanics, where $(q, p) \in \mathbb{R}_{q, p}^{2 n}$, and  
$h \to 0$ is the semiclassical parameter associated with the Planck constant $\hbar$.  
The weak limit $\rho_0 (q, p) = \lim_{h \to 0} \rho_{h} (q, p)$ is non-negatively defined, 
but $\rho_h (q, p)$ itself can be negatively defined over some region 
(the measure of which vanishes in the limit $h \to 0$, see \cite{Karasev_Maslov}). 
It is suggested to perceive $\wt{F} (x; \wt{\alpha})$ as a limit 
\begin{equation} 
\label{eq:limit_F_lambda}
\wt{F} (x; \wt{\alpha}) = \lim_{\lambda \to \infty} \wt{F}_{\lambda} (x; \wt{\alpha}), 
\end{equation}
where $\wt{F}_{\lambda} (x; \wt{\alpha})$ is some function (or, more generally, a distribution), 
depending on a large parameter $\lambda \to \infty$, and the limit is taken in the weak sense. 
It is quite remarkable, that $\wt{F}_{\lambda} (x; \wt{\alpha})$ does \emph{not} need to be non-negatively defined. 
This can lead to an interesting \emph{new physical effect}: the \emph{thermodynamic Bell's inequalities}. 
In quantum mechanics, the violation of Bell's inequalities is related to the fact 
that the probability model in quantum mechanics is different from the probability model in classical mechanics: 
the Wigner's quasiprobability can be negatively defined over some region of the phase space. 
It follows, that once we have $\wt{F}_{\lambda} (x; \wt{\alpha})$ which is negatively defined over 
some region in $\mathbb{R}_{x}^{2 d}$, $x = (x_1, \dots, x_{2 d})$, 
we can mimic the Bell's inequalities and the ``entangled states'' in quasithermodynamics!

Extending the analogy between mechanics and thermodynamics, 
one can look at $\wt{F}_{\lambda} (x; \wt{\alpha})$ along the phenomenological evolution curve 
$\wt{\gamma} = \lbrace \wt{\alpha}^t \rbrace_t \subset \wt{\Lambda}^{\#}$. 
The general shape of the equation for $\wt{F}_{\lambda} (x, \wt{\alpha}^t)$ should then be as follows.  
Equip the affine space $\mathbb{R}_{x}^{2 d}$, $x = (X, J)$, formed by the points 
\begin{equation*} 
x = (X_1, \dots, X_{d}, J_1, \dots, J_d), 
\end{equation*} 
with a symplectic structure $\Omega := \sum_{s = 1}^{d} d J_s \wedge d X_s$, 
where the first $d$ coordinates $(X_1, \dots, X_d)$ correspond to the \emph{extensive} quantities (the ``energies''), and 
the last $d$ coordinates $(J_1, \dots, J_d)$ correspond to the \emph{intensive} quantities (the ``inverse temperatures''). 
Recall, that the volume $V = L^n$ is fixed (for simplicity). 
Consider the \emph{symmetric} Fock space associated with $L^{2} (\mathbb{R}_{x}^{2 d})$: 
\begin{equation*} 
\mathcal{F}^{\#} := \mathbb{C} \oplus L^2 (\mathbb{R}_{x}^{2 d}) \oplus 
 L^2 (\mathbb{R}_{x}^{2 d})^{\otimes_{\mathrm{symm}} 2} \oplus \dots, 
\end{equation*}
where $\otimes_{\mathrm{symm}}$ denotes the symmetric tensor power. 
Let $\wh{\mathcal{K}}_{\lambda} (\wt{\alpha})$, $\wt{\alpha} \in \wt{\Lambda}^{\#}$, 
be a self-adjoint operator on $\mathcal{F}^{\#}$ of the form 
\begin{multline*} 
\wh{\mathcal{K}}_{\lambda} (\wt{\alpha}) = \sum_{m = 0}^{M} 
\frac{\varkappa^{m}}{m!} 
\int dx^{(0)} d x^{(1)} \dots d x^{(m)} 
a^{+} (x^{(0)}) a^{+} (x^{(1)}) \dots a^{+} (x^{(m)}) 
\times \\ \times 
\wh{K}_{\lambda}^{(m)} (\wt{\alpha})
a^{-} (x^{(0)}) a^{-} (x^{(1)}) \dots a^{-} (x^{(m)}),  
\end{multline*}
where 
$x^{(0)}, x^{(1)}, \dots, x^{(m)}$ vary over $\mathbb{R}^{2 d}$, 
$a^{\pm} (x)$ are the bosonic creation and annihilation operators on $\mathcal{F}^{\#}$ \cite{Berezin}, 
\begin{equation*}
[a^{-} (x), a^{-} (x')] = 0 = [a^{+} (x), a^{+} (x')], \quad 
[a^{-} (x), a^{+} (x')] = \delta (x - x'), 
\end{equation*}
$x, x' \in \mathbb{R}^{2 d}$, 
$M$ is a fixed integer, $\varkappa$ is a real parameter (introduced for convenience), 
and the operators $\wh{K}_{\lambda}^{(m)} (\wt{\alpha})$ are given by the commutators with respect to the Moyal product, 
\begin{multline*} 
\wh{K}_{\lambda}^{(m)}  (\wt{\alpha}) = 
\frac{\mathrm{i}}{\lambda^{-1}} 
\bigg\lbrace 
H_{\lambda}^{(m)} \Big(\overset{2}{x^{(0)}} + \frac{\mathrm{i} \lambda^{-1}}{2} J \frac{\overset{1}{\partial}}{\partial x^{(0)} }, 
\dots, 
\overset{2}{x^{(m)}} + \frac{\mathrm{i} \lambda^{-1}}{2} J \frac{\overset{1}{\partial}}{\partial x^{(m)} }; \wt{\alpha} \Big) 
- \\ - 
H_{\lambda}^{(m)} \Big(\overset{2}{x^{(0)}} - \frac{\mathrm{i} \lambda^{-1}}{2} J \frac{\overset{1}{\partial}}{\partial x^{(0)} }, 
\dots, 
\overset{2}{x^{(m)}} - \frac{\mathrm{i} \lambda^{-1}}{2} J \frac{\overset{1}{\partial}}{\partial x^{(m)} }; \wt{\alpha} \Big)  
\bigg\rbrace, 
\end{multline*}
where $J = \| J_{i, j} \|_{i, j = 1}^{2 d}$ is the canonical symplectic matrix, 
$H_{\lambda}^{(m)} (x)$, $m = 0, 1, \dots, M$, are polynomials in $\lambda^{-1}$. 

With this setup one can mimic the constructions of quantum statistical mechanics: 
the quasithermodynamic parameter $\lambda^{-1}$ is similar to the semiclassical parameter $h$, 
and $\varkappa$ is similar to the interaction parameter $g$. 
Observe nonetheless a rather important difference: 
there is a dependence of $\wh{\mathcal{K}}_{\lambda} (\wt{\alpha})$ on the point $\wt{\alpha}$ of the ``huge'' 
Lagrangian manifold $\wt{\Lambda}^{\#}$ described as a subset of pairs $(\wt{F} (\cdot), \wt{\sigma} (\cdot))$ 
by \eqref{eq:Lambda_huge}.
Consider the second quantized analogue of the Wigner's equation for the 
Weyl symbol of the square root of density matrix, 
and break the symmetry in time as in \cite{ZubarevMorozovRopke_vol1}: 
\begin{equation} 
\label{eq:Wigner_second_quantized}
\frac{\partial R_{\lambda, \epsilon}^{t}}{\partial t} + 
\wh{\mathcal{K}}_{\lambda} (\wt{\alpha}^{t}) R_{\lambda, \epsilon}^{t} = 
- \epsilon (R_{\lambda, \epsilon}^{t} - R_{\lambda}^{(\mathit{eq})}), 
\end{equation} 
where $R_{\lambda, \epsilon}^{t} \in \mathcal{F}^{\#}$ is the unknown vector, and 
$R_{\lambda}^{(\mathit{eq})} \in \mathcal{F}^{\#}$ is fixed (the value of $\lim_{\epsilon \to + 0} R_{\lambda, \epsilon}^{t}$ 
corresponding to a state of thermodynamic equilibrium), and $\epsilon = + 0$. 
Then we can perceive the solutions $\wt{F} (x; \wt{\alpha}^{t})$ of the generalized 
Fokker-Planck equation \eqref{eq:Fokker_Planck_phase_space}
as a limit \eqref{eq:limit_F_lambda}, where 
\begin{equation} 
\label{eq:thermodynamic_kinetic}
\wt{F}_{\lambda} (x; \wt{\alpha}^{t}) = \lim_{\epsilon \to +0} 
(R_{\lambda, \varepsilon}^{t}, 
a^{+} (x) a^{-} (x) R_{\lambda, \varepsilon}^{t}),  
\end{equation}
and construct the analogue of BBGKY chain of equations for the higher order correlation functions. 
Furthermore, if there exists a \emph{generating function} of fluctuations, 
\begin{equation*} 
\wt{\mathcal{Z}}_{k_B}^{(\lambda)} (u; \wt{\alpha}^t) := 
\int d x\, \exp  \Big( \frac{\mathrm{i}}{k_B} u x \Big) 
\wt{F}_{\lambda} (x; \wt{\alpha}^{t}), 
\end{equation*}
where $u = (u_1, u_2, \dots, u_{2 d})$ is a parameter varying in a neighbourhood of $\bar 0 = (0, 0, \dots, 0) \in \mathbb{R}^{2 d}$, 
$u x := \sum_{i = 1}^{2 d} u_i x_i$, 
then one can compute the cumulants of fluctuations $\wt{C}_{M}^{(\lambda)} (t)$, 
$M \in \mathbb{Z}_{\geqslant 0}^{2 d} \backslash \lbrace \bar 0 \rbrace$, of the extensive 
thermodynamic quantities $(x_1, \dots, x_d)$ and their flows $(x_{d + 1}, \dots, x_{2 d})$ as follows: 
\begin{equation*} 
\wt{C}_{M}^{(\lambda)} (t) = \Big( - \mathrm{i} k_B \frac{\partial}{\partial u} \Big)^{M} 
k_B \mathrm{ln} 
\wt{\mathcal{Z}}_{k_B}^{(\lambda)} (u; \wt{\alpha}^{t}) \Big|_{u = \bar 0}. 
\end{equation*} 
Note that the function 
$\wt{\mathcal{Z}}_{k_B}^{(\lambda)} ( \mathrm{i} u; \wt{\alpha}^t)$ 
(if the corresponding analytic continuation exists)
restricted to $u_{d + 1} = \dots = u_{2 d} = 0$ 
yields an analogue of a partition function in \emph{nonequilibrium} statistical thermodynamics. 
The asymptotic expansion of $\Phi_{k_B}^{(\lambda)} (u; \wt{\alpha}^{t}) := 
\lambda^{-1} k_B \mathrm{ln} \wt{\mathcal{Z}}_{k_B}^{(\lambda)} ( \mathrm{i} u; \wt{\alpha}^{t})$ in 
$\lambda^{-1} \to 0$ is an analogue of these data in \emph{nonequilibrium} quasithermodynamics. 
The \emph{nonequilibrium} phenomenological thermodynamics 
corresponds to the Lagrangian manifold $\wt{\Lambda}^{\#}$. 
The generalized Fokker-Planck equation is a \emph{phenomenological} equation in this terminology, and 
a link between the ``deformed'' cumulants 
$\lbrace \wt{C}_{M}^{(\lambda)} (t) \rbrace_{M}$ at different moments of time $t$ is induced by \eqref{eq:Wigner_second_quantized}.

\section{Thermocorpuscles}

In the semiclassical approximation of quantum theory it is a common practice 
to denote the small parameter of the asymptotic expansion as $\hbar$. 
In this case it is necessary to keep in mind, that the physical value of the Planck 
constant should not be confused with this parameter. 
For example, one may denote 
$\hbar_{\mathit{phys}} = 6.6262 \times 10^{- 27} \, \mathit{erg}\, \mathit{s}$ 
and after that write $\hbar \to 0$. 
In quasithermodynamics, the small parameter is $\lambda^{-1}$, where $\lambda$ is the rescaling parameter. 
In can be convenient to redenote it as $k_B$ since it stands in the same place as the Boltzmann's constant in the exponent 
linking the free energy and the partition function. 
If we redenote the physical value of the Boltzmann constant, for example, 
as $(k_B)_{\mathit{phys}} = 1.3807 \times 10^{- 16} \, \mathit{erg} \, K^{-1}$, then 
we may write $k_B = \lambda^{-1} \to 0$.

Intuitively, when the Planck's and Boltzmann's constants are introduced into physics, 
$\hbar_{\mathit{phys}}$ is related to the quantization of ``properties'' 
(for example, the energy spectrum), and $(k_B)_{\mathit{phys}}$ is related to the quantization of ``substance''. 
In this sense, the formula for the quantization of energy $E$ of a 1-dimensional 
harmonic oscillator is similar to the ``quantization'' of the number of moles $\nu$, 
\begin{equation*} 
E - E_{0} = \omega \, \hbar_{phys} \, n, \quad 
\nu = R^{-1} \, (k_B)_{\mathit{phys}} \, N, 
\end{equation*}
where $n \in \mathbb{Z}_{\geqslant 0}$ (the number of quanta), 
$N \in \mathbb{Z}_{\geqslant 0}$ (the number of particles), 
$E_{0}$ is the ground level of energy, 
$\omega > 0$ and $R > 0$ are parameters (the frequency of the oscillator and the universal gas constant, respectively). 

In the previous section we have introduced the creation and annihilation operators 
$a^{\pm} (x)$, $x = (X, J)$, where $X = (X_1, \dots, X_d)$ are the extensive thermodynamic quantities 
(like the values of internal energies or the numbers of moles in different parts the system), and 
$J = (J_1, \dots, J_d)$ are the associated flows. 
The volume of the system $V$ is fixed (otherwise, we need to add one more coordinate $x_{2 d + 1}= V$).  
It is natural to perceive these operators as the operators of creation and annihilation operators 
of \emph{``thermoparticles''} (or, another name, could be \emph{``thermocorpuscles''}). 
We work over the space $\mathbb{R}_{X, J}^{2 d}$ rather than $\mathbb{R}_{X}^{d}$, 
or $\mathbb{R}_{J}^{d}$, in order to avoid the 
discussion about the statistics of these thermodynamic particles. 
Are they thermo-fermions, thermo-bosons, or, perhaps, something else? 
Leaving this for another paper, let us look at two other effects. 

\vspace{0.25 true cm}
\emph{Effect no.1: The thermodynamic Bell's inequalities.} 
In \eqref{eq:Fokker_Planck_phase_space} 
we have $\wt{F} (x; \wt{\alpha}^{t})$, $x = (X, J)$, 
which is an analogue (up to a normalization factor $L^n$) of the \emph{classical} probability distribution 
(recall, that $\gamma = \lbrace \wt{\alpha}^{t} \rbrace \subset \wt{\Lambda}^{\#}$ 
defines a curve of evolution of the system).  
In particular, for $\wt{f} (x; \wt{\alpha}) := L^{- n} \wt{F} (x; \wt{\alpha})$, we have 
$\int d x\, \wt{f} (x; \wt{\alpha}^{t}) = 1$, and 
for any test function $\varphi (x) \in C_{0}^{\infty} (\mathbb{R}_{x}^{2 d})$, such that 
$\varphi (x) > 0$, we have 
$\int d x\, \varphi (x) \wt{f} (x; \wt{\alpha}^t) \geqslant 0$. 
On the other hand, for the ``deformed'' distribution 
$\wt{f}_{\lambda} (x; \wt{\alpha}) := L^{- n} \wt{F}_{\lambda} (x; \wt{\alpha})$, one does \emph{not} impose the latter condition, 
and there can exist a test function 
$\varphi_{\lambda} \in C_{0}^{\infty} (\mathbb{R}_{x}^{2 d})$, 
such that $\varphi_{\lambda} (x) > 0$, but  
$\int d x\, \varphi_{\lambda} (x) \wt{F}_{\lambda} (x; \wt{\alpha}^t) < 0$. 
In mechanics, the fact that the Wigner's quasiprobability function can be negatively defined over some region 
of the phase space is responsible for the Bell's inequalities \cite{Bell}. 
In particular, one may introduce a \emph{thermodynamic wave function}, 
considering, for example, a Cauchy problem 
with an initial condition stemming from an element $\psi_{\lambda} (X) \in L^{2} (\mathbb{R}_{X}^{d})$, 
\begin{equation*}
\wt{f} (x; \wt{\alpha}^{t})|_{t = 0} = 
\frac{1}{(2 \pi \lambda^{-1})^{d}}
\int_{\mathbb{R}^{d}} d X' \, 
\exp \Big( - \frac{\mathrm{i}}{\lambda^{-1}} J X' \Big) 
\bar \psi_{\lambda} \Big( X - \frac{X'}{2} \Big) 
\psi_{\lambda} \Big( X + \frac{X'}{2} \Big), 
\end{equation*}
where $X = (x_1, \dots, x_d)$, $J = (J_1, \dots, J_d)$. 
The parameter $\lambda^{-1}$ in this formula is an analogue of the semiclassical parameter $h$ in quantum mechanics. 
If we take, for instance, a WKB-type function 
\begin{equation*} 
\psi_{\lambda} (X) = \exp \Big( \frac{\mathrm{i}}{\lambda^{-1}} S (X) \Big) \varphi_{\lambda} (X), 
\end{equation*} 
where $S (X)$ is \emph{real}, and $\varphi_{\lambda} (X)$ admits an asymptotic expansion 
in the powers of the small parameter $\lambda^{-1}$, 
then we should interpret the derivatives $\partial S (X)/ \partial X_i$, $i = 1, 2, \dots, d$, 
as the flows $J = (J_{1}, J_{2}, \dots, J_{d})$ induced by $X = (X_1, X_2, \dots, X_d)$. 
The limit $\lambda^{-1} \to 0$ corresponds to the nonequilibrium thermodynamic Hamilton-Jacobi equation.  
In other words, the phenomenological thermodynamic flows are the analogues of classical mechanical momenta. 
Intuitively, the leading coefficient $\lim_{\lambda \to \infty} \varphi_{\lambda} (X)$ is something 
similar to a Gaussian exponent concentrated near a point. 
Observe that the quadratic function corresponding to 
the power of this exponent does \emph{not} contain the small parameter $\lambda^{-1}$. 
Considering a thermodynamic system built from several similar subsystems, 
one may construct an \emph{entangled state} and violate the Bell's inequalities. 
The ``observables'' in this case are represented by 
self-adjoint operators on $L^{2} (\mathbb{R}_{X}^{d})$ depending on a parameter $\wt{\alpha} \in \wt{\Lambda}^{\#}$, and 
the inequalities mentioned yield a condition on the fluctuations of  
the measured values of $X_i$ and $J_j$, $i, j = 1, 2, \dots, d$.

\vspace{0.25 true cm}
\emph{Effect no.2: The ``deformed'' Boltzmann's $H$-theorem}. 
 The effect no.1 discussed above corresponds to the first quantization of phenomenological 
nonequilibrium thermodynamics, where the Boltzmann's constant $k_B$ plays, in a certain sense, 
a role similar to the Planck's constant $\hbar$. 
In statistical mechanics the particles interact with each other 
(for example, via an interaction potential). The parameter $g$ in front of 
the interaction potential is sometimes termed the \emph{external Planck constant} $g = \hbar_{\mathit{ext}}$, 
since the commutation relation 
$[\sqrt{g} a^{+} (q), \sqrt{g} a^{-} (q')] = g \delta (q - q')$ for the bosonic 
creation-annihilation operators in configuration space points $q, q' \in \mathbb{R}^{n}$ 
are similar to $[b_{i}^{+}, b_{j}^{-}] = \hbar \delta_{i, j}$, 
where $b_{j}^{\pm} = (q_j \mp \hbar \partial/ \partial q_j)/ \sqrt{2}$, $i, j = 1, 2, \dots, n$. 
In the previous section we have seen, that it is natural to introduce an interaction between the thermocorpuscles 
(the collision integral). 
Let $\wt{F}_{\lambda} (x; \wt{\alpha}^{t})$, $x = (X, J)$, be 
of the shape \eqref{eq:thermodynamic_kinetic} corresponding in the limit $\lambda \to \infty$ to a solution of 
the generalized Fokker-Planck equation \eqref{eq:Fokker_Planck_phase_space}. 
One perceives $\wt{F}_{\lambda} (x; \wt{\alpha}^{t})$ as an analogue of a one-particle 
kinetic function in \emph{quantum} statistical mechanics.  
If $\wt{F}_{\lambda} (x; \wt{\alpha}^{t})$ corresponds to a symbol of a $\lambda^{-1}$-pseudodifferential operator, 
\begin{equation*} 
\wh{F}_{\lambda} (\wt{\alpha}^{t}) := 
\wt{F}_{\lambda} \Big( X, - \mathrm{i} \lambda^{-1} \frac{\partial}{\partial X}; \wt{\alpha}^{t} \Big), 
\end{equation*}
where one uses the Weyl quantization, then it is possible to consider a quantity 
\begin{equation*} 
H_{\lambda} (\wt{\alpha}^{t}) := \mathrm{Tr} \big\lbrace 
\wh{F}_{\lambda} (\wt{\alpha}^{t}) \mathrm{ln} \wh{F}_{\lambda} (\wt{\alpha}^{t})
\big\rbrace. 
\end{equation*}
In analogy with the Boltzmann's theorem, the presence of the collision integral 
implies, that 
\begin{equation*} 
\frac{\partial H_{\lambda} (\wt{\alpha}^{t})}{\partial t} \leqslant 0, 
\end{equation*}
along the nonequilibrium evolution curve $\wt{\gamma} = \lbrace \wt{\alpha}^{t} \rbrace_{t} \subset \wt{\Lambda}^{\#}$. 
This inequality should hold not just in the limit $\lambda \to \infty$, but for \emph{all} values of $\lambda$.

\vspace{0.25 true cm}
As a final remark, it is of interest to point out the following. 
A collection of extensive thermodynamic quantities $X = (X_{1}, X_{2}, \dots, X_{d})$ 
is normally a collection of local internal energies or local densities of chemical substances. 
If we consider the system together with its environment (i.e. the thermostat) as one big system, 
then we have \emph{conservation laws} (for example, the total energy is fixed). 
It follows, that instead of the canonical symplectic structure 
$\omega = \sum_{i = 1}^{d} d J_i \wedge d X_i$ on the whole phase space $\mathbb{R}_{X, J}^{2 d}$, 
one may reduce the problem to a smaller phase space separating the integrals of motion 
(in mechanics this step is termed a reduction of dynamics to a coisotropic submanifold). 
The reduced symplectic structure can be perceived as a kind of \emph{``thermodynamic gauge field''} created by the environment.

\section{Discussion} 

The main motivation of the present paper is a striking (though underappreciated) analogy 
between the roles played by the Planck's and the Boltzmann's constants 
in the mathematical formalism of theoretical physics. 
These two constants were introduced into science at more or less the same time 
(the beginning of the XX-th century) and in a sense capture the \emph{Zeitgeist} (the spirit of the age). 
The concept of a \emph{thermocorpuscle}, i.e. a thermodynamic particle 
with the thermodynamic forces $X = (X_1, X_2, \dots, X_d)$ 
in place of coordinates (internal energy, number of moles, volume, etc.), 
and the associated nonequilibrium thermodynamic flows $J = (J_1, J_2, \dots, J_d)$ in place of momenta, 
is, in a sense, quite natural if one investigates the analogy 
between the Heisenberg's uncertainty principle 
and the Einstein's quasithermodynamic fluctuation theory. 
Informally speaking, the Planck's constant $\hbar$ and the Boltzmann's constant $k_{B}$ 
are very similar concepts, but corresponding to different ``hierarchical'' levels of organization of matter. 

The symmetry between the mechanical and thermodynamic pictures is very much in line with 
the general philosophy of I. Prigogine \cite{Prigogine} who tried to introduce 
the ``irreversibility on a microscopic level of description''. 
A thermocorpuscle is a counterpart of a quantum particle, with $\hbar$ replaced by $k_B$. 
It is of interest to point out, that the analogy between $\hbar$ and $k_B$ is a general 
motivation behind the mathematical constructions in \cite{RuugeFVO1} and \cite{RuugeFVO2}, 
where one describes a kind of \emph{``noncommutative neighbourhood''} around the semiclassical 
and quasithermodynamic parameters. 
Perhaps even more complicated new theoretical structures are needed to capture the nonequilibrium statistical 
physics in all its aspects. 

At this point it is possible to claim that the ``quantization'' of phenomenological thermodynamics 
(with $k_B$ in place of $\hbar$) and the ``second quantization'' of the quantized thermodynamics 
appear to be quite reasonable constructions. 
In particular, conceptually it is important to distinguish between the 
\emph{phenomenological} thermodynamics, \emph{statistical} thermodynamics, and 
the thermodynamics that lives between these two theories, the \emph{quasithermodynamics}. 
This is similar to mechanics: there is \emph{classical} mechanics, \emph{quantum} mechanics, and 
the mechanics existing between the two theories, the \emph{semiclassics} \cite{Maslov_asymp_meth}. 
If we look at the deformation quantization theory \cite{Kontsevich}, then it is a standard practice to denote 
the ``formal'' parameter in the star product as $\hbar$. Now one can see, that $k_B$ is just as good: 
\begin{equation*} 
f (X, J) *_{k_B} g (X, J) = f (X, J) g (X, J) + \sum_{s = 1}^{\infty} k_B^s B_s (f, g) (X, J), 
\end{equation*}
where $f, g \in C^{\infty} (\mathbb{R}_{X, J}^{2 d})$, $B_s$ are bilinear differential operators, $s = 1, 2, 3, \dots$. 

In our days, due to the progress in the field of nanotechnologies,  
one can directly work with systems consisting of relatively small numbers of quantum particles 
(consider microtransistors in electronics, molecular machines in biological physics, etc.). 
For example, the size of a single transistor on a microchip is measured in hundreds of atoms, 
but certainly not in the scale of the Avogadro number. 
The usual empirical laws stemming from phenomenological thermodynamics, 
for example, the linear Ohm's law for the electric current, 
break down at this scale. 
On the other hand, the systems are still too big to be peceived purely quantum mechanically. 
One needs a kind of ``nanothermodynamics'' in this case. 
If we look at mechanics and imagine that we go from a classical description to a quantum description via the semiclassics, 
then one can say that we first construct a Schr\"odinger equation with a semiclassical parameter 
$h \to 0$ in front of partial derivatives, and then formally substitute $h = 1$ 
(this corresponds to the ``real'' Schr\"odinger equation with $\hbar$). 
It looks like a natural speculation, that if one performs a similar step 
in quasithermodynamics (substitute $\lambda^{-1} = 1$), 
then one obtains the ``real'' quantized thermodynamics.


\begin{thebibliography}{99}


\bibitem{Bell}
Bell, J. S.: 
On the problem of hidden variables in quantum mechanics.  
\emph{Rev. Modern Phys.}  \textbf{38} (1966), 447--452.

\bibitem{Berezin}
Berezin, F. A.:
\emph{The method of second quantization.} 
Second edition. Edited and with a preface by M. K. Polivanov. ``Nauka'', Moscow, 1986. 320 pp.


\bibitem{Coutant_Rajeev}
Coutant, A.; S. G. Rajeev: 
Quantum Thermodynamics of Non-Ideal Gases. 
\texttt{arXiv:0807.4632v1 [cond-mat.stat-mech]}




\bibitem{DubnovMaslovNazaikinskii} 
Dubnov, V. L.; Maslov, V. P.; Nazaikinskii, V. E.: 
The complex Lagrangian germ and the canonical operator.  
\emph{Russian J. Math. Phys.}  \textbf{3}  (1995),  no. 2, 141--190.



\bibitem{Fedoryuk}
Fedoryuk, M. V.: 
\emph{The saddle-point method} 
(Russian) 
Izdat. ``Nauka'', Moscow, 1977. 368 pp




\bibitem{Karasev_Maslov}
Karasev, M. V.; Maslov, V. P.:
\emph{Nonlinear Poisson brackets. Geometry and quantization}. 
Translations of Mathematical Monographs, 119. 
American Mathematical Society, Providence, RI, 1993. xii+366 pp. 

\bibitem{Kontsevich}
Kontsevich, M.: 
Deformation quantization of Poisson manifolds, I. 
\texttt{arXiv:q-alg/9709040v1}. 



\bibitem{Maslov_thermo1}
Maslov, V. P.: 
Analytic extension of asymptotic formulas, and the axiomatics of thermodynamics and quasithermodynamics.  (Russian)
\emph{Funktsional. Anal. i Prilozhen.}  \textbf{28}  (1994),  no. 4, 28--41, 95;  
translation in  \emph{Funct. Anal. Appl.}  \textbf{28}  (1994),  no. 4, 247--256 (1995)

\bibitem{Maslov_thermo2}
Maslov, V. P.:
Geometric quantization of thermodynamics, phase transitions and asymptotics at critical points. (Russian) 
\emph{Mat. Zametki} \textbf{56} (1994), no. 3, 155--156; 
translation in \emph{Math. Notes} \textbf{56} (1994), no. 3-4, 984–985 (1995)


\bibitem{Maslov_thermo3}
Maslov, V. P.: 
Geometric "quantization'' of thermodynamics, and statistical corrections at critical points. (Russian) 
\emph{Teoret. Mat. Fiz.} \textbf{101} (1994), no. 3, 433--441; 
translation in \emph{Theoret. and Math. Phys.} \textbf{101} (1994), no. 3, 1466–1472 (1995) 





\bibitem{Maslov_asymp_meth}
Maslov, V. P.: 
\emph{Asymptotic methods and perturbation theory.}
``Nauka'', Moscow, 1988. 311 pp. ISBN: 5-02-013784-7 

\bibitem{Maslov_op_meth}
Maslov, V. P.: 
\emph{Operational methods.} 
Translated from the Russian by V. Golo, N. Kulman and G. Voropaeva. 
Mir Publishers, Moscow, 1976. 559 pp.



\bibitem{Maslov_ultra}
Maslov, V. P.: 
Ultrasecond quantization and ``ghosts'' in quantized entropy. (Russian)  
\emph{Teoret. Mat. Fiz.}  \textbf{129}  (2001),  no. 3, 464--490;  
translation in  \emph{Theoret. and Math. Phys.}  \textbf{129}  (2001),  no. 3, 1694--1716




\bibitem{Maslov_Nazaikinskii} 
Maslov, V. P.; Nazaikinskii, V. E.: 
The tunnel canonical operator in thermodynamics. (Russian)  
\emph{Funktsional. Anal. i Prilozhen.}  \textbf{40}  (2006),  no. 3, 12--29, 96;  
translation in  \emph{Funct. Anal. Appl.}  \textbf{40}  (2006),  no. 3, 173--187




\bibitem{Mori} 
Mori, H.: 
Transport, collective motion, and Brownian motion. 
\emph{Progress of Theoretical Physics} \textbf{33} (1965), no. 3, 423-455



\bibitem{Morozov} 
Morozov, V. G.: 
The generalized Fokker-Planck equation for quantum systems. 
(Russian)  \emph{Teoret. Mat. Fiz.}  \textbf{48}  (1981), no. 3, 373--384.




\bibitem{Onsager}
Onsager, L.: 
Reciprocal relations in irreversible processes. I. 
\emph{Phys. Rev.} \textbf{37} (1931), 405--426 

\bibitem{Prigogine}
Prigogine, I.: 
\emph{Time, Structure and Fluctuations.}
Nobel Lecture, December 8, 1977 




\bibitem{Rajeev1}
Rajeev, S. G.: 
A Hamilton-Jacobi formalism for thermodynamics. 
\emph{Ann. Physics} \textbf{323} (2008), no. 9, 2265–2285.


\bibitem{Rajeev2} 
Rajeev, S. G.: 
Quantization of contact manifolds and thermodynamics.  
\emph{Ann. Physics}  \textbf{323}  (2008),  no. 3, 768--782.













\bibitem{RuugeFVO1} 
Ruuge, A.E.; Van Oystaeyen, F.: 
Distortion of the Poisson bracket by the noncommutative Planck constants. 
\emph{Comm. Math. Phys.} (2010), accepted. 

\bibitem{RuugeFVO2} 
Ruuge, A.E.; Van Oystaeyen, F.: 
$q$-Legendre transformation: partition functions and quantization of the Boltzmann constant. 
\emph{J. Phys. A: Math. Theor.} \textbf{43} (2010), 345203 (30pp)



\bibitem{ZubarevMorozovRopke_vol1}
Zubarev, D.; Morozov, V.; R\"opke, G.: 
\emph{Statistical mechanics of nonequilibrium processes. Vol. 1: Basic concepts, kinetic theory.} 
John Wiley \& Sons, 1996. 270 pp. 


\bibitem{ZubarevMorozovRopke_vol2} 
Zubarev, D.; Morozov, V.; R\"opke, G.: 
\emph{Statistical Mechanics of Nonequilibrium Processes, Vol. 2: Relaxation and hydrodynamic processes} 
John Wiley \& Sons, 1997. 375 pp. 








\bibitem{Zwanzig} 
Zwanzig, R.: 
Memory effects in irreversible thermodynamics. 
\emph{Phys. Rev.} \textbf{124} (1961), no. 4, 983--992





\end{thebibliography}
\end{document}